\newif\ifjournal  
\newif\ifblind    

\ifjournal
  \ifblind
    \documentclass[sigconf,review,anonymous]{acmart}
  \else
    \documentclass[sigconf]{acmart}
  \fi
  \usepackage{url}

  \copyrightyear{2023}
  \acmYear{2023}
  \setcopyright{licensedusgovmixed}\acmConference[ICPP 2023]{52nd International Conference on Parallel Processing}{August 7--10, 2023}{Salt Lake City, UT, USA}
  \acmBooktitle{52nd International Conference on Parallel Processing (ICPP 2023), August 7--10, 2023, Salt Lake City, UT, USA}
  \acmPrice{15.00}
  \acmDOI{10.1145/3605573.3605594}
  \acmISBN{979-8-4007-0843-5/23/08}

  \newcommand\ttl[1]{\texttt{#1}}
\else
  \documentclass[twoside,leqno,twocolumn]{article}
  \usepackage{orcidlink}

  \usepackage[sf={lining,proportional},tt={variable,lining,tabular},rm={lining,proportional}]{cfr-lm}
  \newcommand\ttl[1]{\texttt{\textlg{#1}}}
\fi

\usepackage{algorithm}
\usepackage[noend]{algpseudocode}
\usepackage{amsfonts}
\usepackage{booktabs}
\usepackage{breakcites}
\usepackage[noabbrev]{cleveref}
\usepackage{standalone}
\usepackage[caption=false,font=footnotesize,labelfont=sf,textfont=sf]{subfig}
\usepackage{multirow}
\usepackage{tikz}
\usetikzlibrary{positioning, shadows, arrows, backgrounds, scopes}
\usepackage{xcolor}
\usepackage{xspace}

\makeatletter
\def\blfootnote{\xdef\@thefnmark{}\@footnotetext}
\makeatother

\newcommand{\dbscan}{\textsc{Dbscan}\xspace}
\newcommand{\dbscans}{\textsc{Dbscan*}\xspace}
\newcommand{\fdbscan}{\textsc{FDbscan}\xspace}
\newcommand{\fdbscandense}{\textsc{FDbscan-DenseBox}\xspace}
\newcommand{\pdsdbscan}{\textsc{Pdsdbscan-S}\xspace}
\newcommand{\tepp}{\textsc{Tepp}\xspace}
\newcommand{\gdbscan}{\textsc{G-Dbscan}\xspace}

\newcommand{\unionfind}{\textsc{Union-Find}\xspace}
\newcommand{\union}{\textsc{Union}\xspace}
\newcommand{\find}{\textsc{Find}\xspace}

\newcommand{\hdbscan}{\textsc{Hdbscan*}\xspace}
\newcommand{\kdtree}{\textit{k}-d tree\xspace}
\newcommand{\minpts}{\textit{minPts}\xspace}
\newcommand{\eps}{\varepsilon}

\newcommand{\dataset}[1]{\emph{#1}\xspace}

\newcommand{\nvidiagpu}{Nvidia A100\xspace}
\newcommand{\amdgpu}{AMD MI250X\xspace}
\newcommand{\amdcpu}{AMD EPYC 7763\xspace}

\begin{document}

\title{Fast tree-based algorithms for DBSCAN for low-dimensional data on GPUs}

\ifjournal
  \unless\ifblind
    \author{Andrey Prokopenko}
    \orcid{0000-0003-3616-5504}
    \affiliation{%
      \institution{Oak Ridge National Laboratory}
      \streetaddress{1 Bethel Valley Rd}
      \city{Oak Ridge}
      \state{Tennessee}
      \country{USA}
      \postcode{37830}
    }
    \email{prokopenkoav@ornl.gov}

    \author{Damien Lebrun-Grandi\'e}
    \orcid{0000-0003-1952-7219}
    \affiliation{%
      \institution{Oak Ridge National Laboratory}
      \streetaddress{1 Bethel Valley Rd}
      \city{Oak Ridge}
      \state{Tennessee}
      \country{USA}
      \postcode{37830}
    }
    \email{lebrungrandt@ornl.gov}

    \author{Daniel Arndt}
    \orcid{0000-0001-8773-4901}
    \affiliation{%
      \institution{Oak Ridge National Laboratory}
      \streetaddress{1 Bethel Valley Rd}
      \city{Oak Ridge}
      \state{Tennessee}
      \country{USA}
      \postcode{37830}
    }
    \email{arndtd@ornl.gov}

    \renewcommand{\shortauthors}{Prokopenko et al.}
  \fi
\else
  \author{
    A.~Prokopenko\thanks{Oak Ridge National Laboratory}\enskip\orcidlink{0000-0003-3616-5504},
    D.~Lebrun-Grandi\'e\footnotemark[1]\enskip\orcidlink{0000-0003-1952-7219},
    D.~Arndt\footnotemark[1]\enskip\orcidlink{0000-0001-8773-4901}
  }
  \date{}
\fi

\ifjournal
  \begin{abstract}
    DBSCAN is a well-known density-based clustering algorithm to discover arbitrary
shape clusters. While conceptually simple in serial, the algorithm is
challenging to efficiently parallelize on manycore GPU architectures. Common
pitfalls, such as asynchronous range query calls, result in high thread
execution divergence in many implementations. In this paper, we propose a new
framework for GPU-accelerated DBSCAN, and describe two tree-based algorithms within
that framework. Both algorithms fuse the search for neighbors with updating
cluster information, but differ in their treatment of dense regions of the
data. We show that the time taken to compute clusters is at most twice that of
determination of the neighbors. We compare the proposed algorithms with existing
CPU and GPU implementations, and demonstrate their competitiveness and
performance using a fast traversal structure (bounding volume
hierarchy) for low dimensional data. We also show that the memory usage can be
reduced by processing object neighbors dynamically without storing them.

  \end{abstract}

\fi

\maketitle

\unless\ifjournal
  DBSCAN is a well-known density-based clustering algorithm to discover arbitrary
shape clusters. While conceptually simple in serial, the algorithm is
challenging to efficiently parallelize on manycore GPU architectures. Common
pitfalls, such as asynchronous range query calls, result in high thread
execution divergence in many implementations. In this paper, we propose a new
framework for GPU-accelerated DBSCAN, and describe two tree-based algorithms within
that framework. Both algorithms fuse the search for neighbors with updating
cluster information, but differ in their treatment of dense regions of the
data. We show that the time taken to compute clusters is at most twice that of
determination of the neighbors. We compare the proposed algorithms with existing
CPU and GPU implementations, and demonstrate their competitiveness and
performance using a fast traversal structure (bounding volume
hierarchy) for low dimensional data. We also show that the memory usage can be
reduced by processing object neighbors dynamically without storing them.

\fi

\unless\ifblind
  \blfootnote {%
  This manuscript has been authored by UT-Battelle, LLC, under contract
  DE-AC05-00OR22725 with the U.S. Department of Energy. The United States
  Government retains and the publisher, by accepting the article for publication,
  acknowledges that the United States Government retains a nonexclusive, paid-up,
  irrevocable, world-wide license to publish or reproduce the published form of
  this manuscript, or allow others to do so, for United States Government
  purposes.
  }
\fi
\section{Introduction}\label{s:introduction}
Clustering is a data mining technique that splits a set of objects into
disjoint classes (\emph{clusters}), each containing similar objects.
\dbscan (Density-Based Spatial Clustering of Applications with
Noise)~\cite{ester1996} is a density-based clustering algorithm. It is
useful when the number of clusters or their shape is not known \emph{a priori}.
It is used in a diverse set of applications such as
\iffalse
bioinformatics~\cite{madeira2004}, noise filtering and outlier
detection~\cite{birant2007}, halos identification in
cosmology~\cite{sewell2015}, image segmentation~\cite{shen2016}, and others.
\else
bioinformatics, noise filtering and outlier
detection, cosmology, image segmentation, and others.
\fi

The \dbscan algorithm requires the identification of close neighbors
for each data point. Its breadth-first search nature makes parallelization a
challenge. Major progress occurred in the last two decades, starting from the
master-slave~\cite{xu1999, arlia2001} and MapReduce~\cite{he2011,dai2012}
approaches, and transitioning to using shared memory~\cite{patwary2012,
patwary2015, gotz2015, kumari2017, wang2020} and GPU~\cite{bohm2009, thapa2010,
andrade2013, welton2013, welton2014, loh2015, gowanlock2017,
gowanlock2019,gowanlock2021,nagarajan2023} implementations, and even approximate
algorithms~\cite{patwary2014, gan2017, lulli2016, chen2019}. Using the \unionfind
technique for cluster labeling, introduced in~\cite{patwary2012}, was a
particularly important breakthrough as it fundamentally changed the nature of
the algorithm, breaking with its breadth-first search origins.

In this work, we first introduce a general parallel algorithm with sufficient
degree of parallelism for thousands of cores available on GPUs. All components
of the algorithm are executed on a GPU.

We then propose two concrete implementations. We prioritize using an indexing
structure with a fast batched neighborhood search to maintain algorithm
performance. Specifically, we use a bounding volume
hierarchy (BVH), a structure predominantly used in computer graphics for ray
tracing~\cite{meister2021}. We combine it with a synchronization-free
union-find technique introduced in ~\cite{jaiganesh2018}. Our approach allows
processing the found neighboring points on-the-fly, reducing the overall memory
consumption of the limited GPU memory. We introduce several traversal
optimization techniques and reduce the number of distance calculations used by
the algorithm in dense regions. We show significant performance
improvements over available multi-threaded CPU and GPU \dbscan implementations. Since the
local \dbscan implementation is an inherent component of a full distributed
algorithm, the proposed algorithm can be easily plugged into most distributed
frameworks to improve the overall performance.

This paper focuses on the low-dimensional (e.g., spatial) data for two
reasons. First, this work was motivated by scientific simulations, such as
cosmology. The data in these simulation is commonly low-dimensional (e.g., 3D),
and the main challenge lies in its size, reaching 500 million data points for a
single GPU (with a full simulation requiring hundreds or thousands of GPUs).
Given that the data is often analyzed \emph{in-situ}, it is imperative for the
underlying algorithm to be fast. Second, an implementation of a tree-based indexing structure
for high dimensions on an accelerator such as GPU is a challenging task in
itself, as the ``curse of dimensionality'' creates challenges for the popular
data structures used for low-dimensional data~\cite{bohm2001}.

Our key contributions are:
\begin{itemize}
  \item
    We reformulate the \dbscan algorithm to expose more parallelism required
    for an efficient GPU implementation.
  \item
    We use BVH as the search index, selected for its high efficiency on GPUs.
  \item
    We develop a new way to reduce the number of calculations in the dense data
    regions through including dense cells into a hybrid BVH hierarchy together
    with sparse data, combining the benefits of both search index and
    grid-based methods.
  \item
    We provide the first performance portable algorithm and implementation for
    the \dbscan, and provide a comprehensive set of experiments on three
    architectures (\amdcpu CPU, \nvidiagpu GPU, \amdgpu GPU).
\end{itemize}

The remainder of the paper is organized as follows. \Cref{s:background}
introduces the \dbscan algorithm and related work. \Cref{s:algorithm} describes
a general framework for a GPU \dbscan implementation allowing for fine-grained
parallelism. In \Cref{s:fdbscan}, we describe two tree-based algorithms within
that framework. Finally, we demonstrate the algorithm performance and
performance portability in \Cref{s:results} and derive our conclusions and
future work in \Cref{s:conclusions}.

\section{Background}\label{s:background}

\subsection{DBSCAN algorithm}\label{s:dbscan}

We briefly outline the \dbscan algorithm in this Section, referring the readers
to~\cite{ester1996} for more details.

Let $X$ be a set of $n$ points to be clustered. For a point to be in a cluster,
the density in its neighborhood has to exceed some threshold, i.e., its
neighborhood has to contain at least a minimum number of points.
This is formalized using two user-provided parameters:
$\minpts \in \mathbb{N}^+$ and $\eps \in \mathbb{R}^+$.

An \emph{$\eps$-neighborhood} of a point $x$ is defined as $N_\eps(x)
= \bigl\{y \in X \, | \allowbreak \, dist(x, y) \le \eps\bigr\}$, with $dist(\cdot,\cdot)$ being
a distance metric for the set $X$ (e.g., Euclidean). The $\minpts$ parameter
defines the minimum number of points for a point to be considered
inside a cluster, and a point $x$ is called a \emph{core point} if $|N_\eps(x)|
\ge \minpts$. A point $y$ is \emph{directly density-reachable} from a point
$x$ if $x$ is a core point and $y \in N_\eps(x)$. A point $y$ is
\emph{density-reachable} from a point $x$ if there is a chain of points $x_1,
\dots, x_n$, $x_1 = x$, $x_n = y$, such that $x_{i+1}$ is directly
density-reachable from $x_i$. Points $x$ and $y$ are called
\emph{density-connected} if there exists a point $z$ in $X$ such that both $x$ and $y$
are density-reachable from $z$. Finally, a point $x$ is called a \emph{border
point} if it is density-reachable from a core point, but is not a core point
itself. The points that are not core or border points are called \emph{noise}
and are considered to be outliers not belonging to any cluster.
Any cluster then consists of a combination of core points (at least one) and
border points (possibly, none). Note, that as a border point may be density-reachable
from multiple core points, it could potentially belong to multiple
clusters. Implementations of the algorithm may differ in their handling of such
border points, but typically assign them to a single cluster.

\begin{figure}[t]
\captionsetup{labelformat=empty}
\begin{algorithm}[H]
\caption{\dbscan algorithm}\label{a:dbscan}
\begin{algorithmic}[1]
\small
\Procedure{Dbscan}{$X, \minpts, \eps$}
\For {each unvisited point $x \in X$}
    \State mark $x$ as visited
    \State $N \gets \ttl{GetNeighbors}(x, \eps)$  \label{l:dbscan:N}
    \If {$|N| < \minpts$}
        \State mark $x$ as noise                    \label{l:dbscan:noise}
    \Else
        \State {$C \gets \{x\}$}
        \ForAll {$y \in N$}
            \State $N \gets N \backslash y$
            \If {$y$ is not visited}
                \State mark $y$ as visited
                \State $\bar{N} \gets \ttl{GetNeighbors}(y, \eps)$
                \If {$|\bar{N}| \ge \minpts$}
                    \State $N \gets N \cup \bar{N}$
                \EndIf
            \EndIf
            \If {$y$ is not a member of any cluster}
                \State $C \gets C \cup \{y\}$
            \EndIf
        \EndFor
    \EndIf
\EndFor
\EndProcedure
\end{algorithmic}
\end{algorithm}
\end{figure}

The special case of $\minpts = 2$ (sometimes called Friends-of-Friends in the
cosmology literature) is equivalent to finding strongly
connected components in the adjacency graph $G = (V, E)$, where $V = X$ and two
vertices $x$ and $y$ have an (undirected) edge between them if $dist(x, y) \le
\eps$. In this case, there are no border points, and a point either belongs
to a cluster as a core point, or is in the noise.

The pseudocode for the \dbscan algorithm is shown in the \Cref{a:dbscan}. The
algorithm starts at an arbitrary point $x \in X$, computing its
$\eps$-neighborhood $N$ (\cref{l:dbscan:N}). If $x$ is not a core point, i.e.
$|N| < \minpts$, $x$ is tentatively marked as noise
(\cref{l:dbscan:noise}), and another point is chosen. Otherwise, the algorithm
constructs a new cluster $C$ by incrementally adding points that are density-reachable from $x$
in a breadth-first search manner (lines 8-17), including the points that
may have been previously marked as noise. Border points are assigned
to the first encountered cluster that they are density-reachable from.
The algorithm has a computational complexity of $O(n^2)$, or $O(n \log n)$ if a
spatial indexing structure (e.g., k-d tree~\cite{bentley1975} or
R-tree~\cite{guttman1984}) is used.

\dbscans proposed in~\cite{campello2013} simplified the algorithm by removing
the notion of border points completely, thereby improving consistency with the
statistical interpretation of clustering. While not addressed in this work, the
algorithms proposed in this paper can be easily adapted for \dbscans, with
several further optimizations possible.

\subsection{Related work}\label{s:related_gpu_work}

Many papers detail parallelization techniques in distributed~\cite{xu1999,
he2011, patwary2012, welton2013, welton2014, patwary2015, gotz2015, hu2017} and
shared memory~\cite{patwary2012, patwary2015, kumari2017, wang2020} contexts. Here,
we focus on the works addressing the algorithm parallelization using GPUs.

\cite{bohm2009} proposed two algorithms. CUDA-DClust creates
sub-clusters (chains) of points density-reachable from each other. Multiple
chains are created simultaneously in parallel on a GPU. The algorithm keeps track
of chain collisions through a collision matrix, which is resolved on the CPU in the
final stage. CUDA-DClust* is an extension of CUDA-DClust that uses an indexing
technique (based on a constant number of directory level partitions) for
the computation of $N_\eps(x)$.
Two slight modifications of CUDA-DClust, reducing the number of memory
transfers between a CPU and a GPU, and identifying core points prior to cluster
generation, were proposed in Mr. Scan~\cite{welton2013}.
\cite{thapa2010} offloads the $N_\eps(x)$ computation to the GPU by assigning points in
$X$ to different threads, which check the distance to $x$ in parallel.
G-DBSCAN~\cite{andrade2013} constructs the adjacency graph using an all-to-all
computation on the GPU, and then executes a parallel breadth-first search with level
synchronization.
An extension of CUDA-DClust is realized in CudaSCAN~\cite{loh2015}, which trims
the amount of required distance evaluations by partitioning a data set
into subregions and performing local clustering within the sub-regions in
parallel.
A special case of \dbscan with $\minpts = 2$ was studied in~\cite{sewell2015},
where an implicit graph structure combined with a disjoint-set algorithm  was
used to find strongly connected components utilizing a cell partitioning of the
domain as an indexing structure.
\cite{gowanlock2017} utilizes a hybrid CPU-GPU approach in which the neighbors
of each point are first identified on the GPU, then the neighbor list is transferred to
the host, where the clustering is performed.
In~\cite{mustafa2019}, the authors compared existing GPU implementations
(the algorithm in~\cite{thapa2010}, CUDA-DClust*~\cite{bohm2009} and G-DBSCAN~\cite{andrade2013}),
and found G-DBSCAN to be the fastest but requiring significantly more memory
($166\times$ of CUDA-DClust) due to storing the adjacency graph.
\cite{gowanlock2019a} extended the work~\cite{gowanlock2017}, addressing the
limitations of the GPU memory by using a batched mode to incrementally compute
$N_\eps(x)$, and explored avoiding distance calculations in the dense regions
by superimposing a regular grid over the domain, with a special treatment of
the cells containing at least \minpts{} points, called \emph{dense cells}.
CUDA-DClust+ of~\cite{gowanlock2021} further improved CUDA-DClust by reducing the
amount of CPU-GPU communications and moving more kernels to GPU.
A new approach to implement DBSCAN using Nvidia RTX (ray-tracing hardware) was
proposed in~\cite{nagarajan2023}, improving performance for low-density datasets.

This work shares similarities with several of the mentioned algorithms.
Similar to~\cite{andrade2013}, our algorithm operates on the adjacency
graph. However, in this work, the graph is implicit and is never fully formed,
resolving many of the memory constraints of the algorithm identified
in~\cite{mustafa2019}.
Compared to~\cite{sewell2015}, which can be seen as a precursor, this work
implements the full DBSCAN algorithm, uses a synchronization-free non-iterative
union-find algorithm, and uses and optimizes a tree-based different indexing
structure.
Like in this work, \cite{gowanlock2017} identified batched neighbor search as a
key to performance; however, that approach produced a full adjacency graph and
relied on CPU for the clustering itself.
We follow the ideas introduced in~\cite{welton2013,sewell2015,song2018,gowanlock2019a},
and utilize an auxiliary regular grid to reduce the number of distance
calculations. Compared to the mentioned works, however, the cells of the grid
become primitives used in the construction of the tree, both reducing the size
of the tree, and allowing for an easier merge of dense cells.
Finally, compared to most of the works mentioned, the algorithm only uses the GPU
with no support from a CPU, requiring no data transfer between host and device
memories during the execution.

\section{Parallel DBSCAN framework for GPUs}\label{s:algorithm}

\subsection{Disjoint-set based DBSCAN}\label{s:dbscan_union_find}

The main obstacle to the parallelization of the \dbscan algorithm in the
original form (\Cref{a:dbscan}) is its breadth-first manner of encountering new
points, and the linear time required to update the existing neighbor set $N$.
The algorithm proposed in~\cite{patwary2012} breaks with its breadth-first
nature, and serves as the foundation for this work. Instead of maintaining an
explicit list of indices, the authors used the \unionfind~\cite{tarjan1979}
approach to maintain a disjoint-set data structure. The approach relies on two
main operations: \union and \find. $\find(x)$ determines the representative of a set that
a point $x$ belongs to, while $\union(x,y)$ combines the sets that $x$ and $y$
belong to.

The \unionfind algorithm is typically implemented using trees. For any point
$x$, its representative, returned by $\find(x)$, is the root of the tree containing $x$. The $\union(x,y)$
operation merges two trees (containing $x$ and $y$) by pointing the parent
pointer of one tree root (e.g., $\find(x)$) to the other ($\find(y)$). If $x$
and $y$ belong to the same set, then $\find(x)$ and $\find(y)$ return the same
index, and no merging is required.  The procedure starts with creating a forest
of singleton non-overlapping trees, each corresponding to a set consisting of a
single data point. The method proceeds by progressively combining pairs of sets
through merging corresponding trees.

From an implementation perspective, the trees in the \unionfind algorithm are
stored using a flat array, which we will refer to as \emph{labels}. A parent of
a node in a tree is then the value of the label corresponding to that node. The
\find operation follows the values of labels until encountering an index
that is the same as its label, which indicates that it is the root of that tree.
Two trees are merged by changing the label of the root of one of the
trees to that of the other.

\begin{figure}[t]
\captionsetup{labelformat=empty}
\begin{algorithm}[H]
\caption{Disjoint-set \dbscan algorithm}\label{a:dbscan_union_find}
\begin{algorithmic}[1]
\small
\Procedure{DSDbscan}{$X, \minpts, \eps$}
\For {each point $x \in X$}
    \State $N \gets \ttl{GetNeighbors}(x, \eps)$    \label{l:dbscan_union_find:neigh}
    \If {$|N| \ge \minpts$}
        \State mark $x$ as core point
        \For {each $y \in N$}
            \If {$y$ is marked as a core point}     \label{l:dbscan_union_find:core_check}
                \State $\ttl{Union}(x, y)$  \label{l:dbscan_union_find:union1}
            \ElsIf {$y$ is not a member of any cluster}
                \State mark $y$ as a member of a cluster
                \State $\ttl{Union}(x,y)$   \label{l:dbscan_union_find:union2}
            \EndIf
        \EndFor
    \EndIf
\EndFor
\EndProcedure
\end{algorithmic}
\end{algorithm}
\end{figure}

\Cref{a:dbscan_union_find} reproduces the disjoint-set
\dbscan (\textsc{DSDbscan}) algorithm as proposed in~\cite{patwary2012}
(Algorithm 2), shown here for completeness. Each point now only computes its
own neighborhood (\Cref{l:dbscan_union_find:neigh}). If it is a core point, its
neighbors are assigned to the same cluster
(\Cref{l:dbscan_union_find:union1,l:dbscan_union_find:union2}).

In the original paper, a thread or an MPI rank executed the algorithm
sequentially for a subset of data constructed by partitioning, and merged the
results in parallel to obtain the final clusters. For GPUs, however, more
available parallelism is desired to improve the efficiency. In the next Section,
we reformulate the algorithm to allow that.

\subsection{Parallel disjoin-set based DBSCAN}\label{s:dbscan_union_find_parallel}

While the amount of the parallelism in \Cref{a:dbscan_union_find} may be
sufficient for shared- or distributed-memory implementations, it is insufficient
for GPU implementations with thousands or tens of thousands threads. Therefore,
our goals were to reformulate the algorithm to accommodate such a high number of
threads, and to to reduce thread execution divergence (executing different
code) and data divergence (reading or writing disparate locations in memory) in
the algorithm.

\Cref{a:dbscan_union_find} consists of two distinct kernels: the neighbor
search, and the disjoint-set structure update. It is clear that the former is
more computationally demanding than the latter. Without taking appropriate
care, calling \ttl{GetNeighbors} asynchronously by different threads will
result in high execution and data divergence. This is especially true when an
index structure, such as \kdtree or R-* tree, is used. Thus, the neighbor
searches are executed simultaneously for all points a batched mode.

We next address the limited amount of available GPU memory.
Storing all the neighbors found on Line 3 for all threads executed at the same
time may not be possible, given that the number of such neighbors may be a
significant fraction of the overall dataset size. This can be addressed by
observing that the neighbor list is being used in two different contexts. For
assessing whether a point is a core point on Line 4, the only
information required is the number of neighbors, but not the neighbors
themselves. In the loop on Line 6, the neighbors are assigned to the same cluster
as part of the \unionfind algorithm. The key observation here is that the neighbors may be
processed independently and in any order. In other words, it is possible to
process them as they are determined and execute the \union operation on-the-fly
for each neighbor, discarding the found neighbor after that.

Given these findings, we split the algorithm into two
phases. In the first phase, called \emph{preprocessing}, the algorithm
determines the core points. We note that while it is possible to do this by
computing the exact number of neighbors $|N_\eps(x)|$, it is not necessary. If the neighbors
of a point are discovered incrementally (whether through a tree traversal, or
otherwise), it is sufficient to encounter just \minpts{} neighbors to determine a
core point (unless executing a sweep over multiple values of \minpts{}).

The second phase, called \emph{main}, proceeds with the knowledge of core
points, and executes $\union(x,y)$ for each pair of close neighbors as they are
being discovered. This general formulation leaves a lot of room for
optimizations. For example, many of the distance calculations may be
eliminated. We examine this in more detail in \Cref{s:fdbscan}.

The two-phase approach results in dramatic reduction of the consumed memory and
in better avoidance of thread and data divergence. The memory consumption does
not depend on the values of $\eps$ and $\minpts$ and is linear with respect to
the number of points in a dataset (assuming the used search index obeys this,
too). This makes it possible to execute the algorithm for much larger datasets.
As was observed in earlier works, algorithms that store full neighbor lists
(e.g., G-DBSCAN) tend to run out of memory even for smaller datasets,
particularly in situations where $|N_\eps(x)| \gg \minpts$ for a significant
fraction of points.

An additional advantage of the two-phase approach is that it exposes edge-level
parallelism in addition to the vertex-level parallelism. One could consider
using multiple threads collaborating on a single point, with each thread
assigned one of the outgoing edges in the adjacency graph. Such an approach
would require implementing a search index (tree or otherwise) with
multiple threads collaborating on a single search query.

\begin{figure}[t]
\captionsetup{labelformat=empty}
\begin{algorithm}[H]
\caption{Parallel disjoint-set \dbscan algorithm\label{a:fdbscan}}
\begin{algorithmic}[1]
\small
\Procedure{PDSDbscan}{$X, \minpts, \eps$}
\If {$\minpts > 2$}
  \For {each point $x \in X$ \textbf{in parallel}}
      \State determine whether $x$ is a core point    \label{l:fdbscan:is_core_point}
  \EndFor
\EndIf
\For {each pair of points $x, y$ such that $dist(x,y) \le eps$ \textbf{in parallel}} \label{l:fdbscan:edges}
  \If {$x$ is a core point}
    \If {$y$ is a core point}
      \State {$\ttl{Union}(x, y)$}
    \ElsIf {$y$ is not yet a member of any cluster}
      \State \textbf{critical section:}
      \State {\hskip1.5em mark $y$ as a member of a cluster}
      \State {\hskip1.5em $\ttl{Union}(x, y)$}
    \EndIf
  \EndIf
\EndFor
\EndProcedure
\end{algorithmic}
\end{algorithm}
\end{figure}

The pseudocode for the parallel disjoint-set \dbscan (\textsc{PDSDbscan}) algorithm
is shown in \Cref{a:fdbscan}. The preprocessing phase is executed on Lines 3-4.
The check on Line 2 allows the preprocessing phase to be skipped in the special case
when $\minpts = 2$. In this case, any pair of points found within distance
$\eps$ in the main phase is guaranteed to consist of core points. The \unionfind algorithm is performed
on Lines 8 and 11.

The operations on Lines 11 and 12 must be executed
in a single critical section. If a thread is marking $y$ as a member
of its own cluster, no other thread is allowed to execute \union with $y$.
Otherwise, it may lead to the ``bridging'' effect, where a border point within
distance $\eps$ of two separate clusters may result in merging those clusters
together. In practice, it is possible to use the labels array for both
clustering information, and as an indicator for whether a border point is a
member of a cluster. In this approach, the check on Line 9 compares the
label of point $y$ with $y$. If they are identical, the label is assigned the
representative of $x$. It allows us to replace the critical section with a single
atomic compare-and-swap operation.

In summary, the proposed approach allows execution of the full \dbscan
algorithm on a GPU fully in parallel. No data transfers between a CPU and a GPU
are necessary as long as both the data and the chosen search index fit into the
GPU memory.

\section{Tree-based algorithms}\label{s:fdbscan}

\subsection{FDBSCAN}

\fdbscan (``fused'' DBSCAN) fuses tree traversal with the \unionfind algorithm.
It uses a bounding volume hierarchy (BVH), a structure commonly used in
computer graphics for ray tracing, for the search index. While any tree can be
used, BVH has been shown to be very efficient for low-dimensional data on GPUs.
Linear BVH (LBVH) (e.g., ~\cite{karras2012}), are well suited for GPUs, with
low data and thread divergence during both construction and traversal.

The parallelization is done over all points of a dataset, with each thread
assigned a single point. The neighbor search is executed in bulk (i.e., with
all threads launching at the same time). The threads are sorted using
space-filling curve to reduce data and execution divergence during the
traversal. Each thread executes a stack-less top-down traversal. In the
preprocessing phase, we use the recommendation from the previous Section,
terminating the traversal of a thread once a $\minpts$ neighbors are
encountered. In the main phase, the algorithm executes \union operation when a
new neighbor is found, without storing said neighbor.

\begin{figure}[t]
  \centering
\definecolor{ColorInternalIndex}{HTML}{E4C6B3}
\definecolor{ColorLeafIndex}{HTML}{C0C7B8}
\definecolor{ColorSplit}{HTML}{D48C8B}
\definecolor{ColorNode}{HTML}{ECECEC}
\definecolor{ColorGray}{HTML}{666666}
\definecolor{ColorSentinel}{HTML}{DBDB8D}
\definecolor{ColorRope}{HTML}{AEC7E8}

\begin{tikzpicture}[
    box/.style={rectangle, draw=black, fill=ColorNode, text=ColorNode, text centered, minimum height=0.3cm},
    v_box/.style={rectangle, draw=black, fill=ColorNode, text=ColorGray, align=center},
    internal_parent/.style={circle, draw=black, fill=ColorSplit, text=ColorSplit},
    internal_node/.style={circle, draw=black, fill=ColorInternalIndex, text=black},
    gray_internal_node/.style={circle, draw=black, fill=ColorGray!30, text=black},
    leaf_node/.style={circle, draw=black, fill=ColorLeafIndex, text=black},
    gray_leaf_node/.style={circle, draw=black, fill=ColorGray!30, text=black},
    sentinel_node/.style={circle, draw=black, fill=ColorSentinel, text=black},
    text centered,
    font=\bf,
    anchor=north west,
    level distance=0.5cm,
    growth parent anchor=south,
    line width=1pt,
    >=stealth,
    tips=proper
  ]
  {[on background layer]
    \node(box_0)[box, minimum width=11.0cm, opacity=0] at (0.0, 0){};
    \node(box_3)[box, minimum width= 5.0cm, opacity=0] at (0.0,-2){};
    \node(box_4)[box, minimum width= 5.0cm, opacity=0] at (6.0,-1){};
    \node(box_1)[box, minimum width= 2.0cm, opacity=0] at (0.0,-3){};
    \node(box_2)[box, minimum width= 2.0cm, opacity=0] at (3.0,-3){};
    \node(box_5)[box, minimum width= 3.5cm, opacity=0] at (7.5,-2){};
    \node(box_6)[box, minimum width= 2.0cm, opacity=0] at (7.5,-3){};
    \fill[fill=gray!30] (-1.5,-5)--(7.8,-5)--(2.8, 0.0);
  }
  \node(internal_node_0)[internal_node, left =-0.6cm of box_0, opacity=0] {0};
  \node(internal_node_3)[internal_node, right=-0.6cm of box_3, opacity=0] {3};
  \node(internal_node_4)[internal_node, left =-0.6cm of box_4, opacity=0] {4};
  \node(internal_node_1)[internal_node, right=-0.6cm of box_1, opacity=0] {1};
  \node(internal_node_2)[internal_node, left =-0.6cm of box_2, opacity=0] {2};
  \node(internal_node_5)[internal_node, left =-0.6cm of box_5, opacity=0] {5};
  \node(internal_node_6)[internal_node, right=-0.6cm of box_6, opacity=0] {6};
  \node(internal_parent_0)[internal_node, right=4.75cm of internal_node_0] {0};
  \node(internal_parent_3)[gray_internal_node, left=1.75cm of internal_node_3] {3};
  \node(internal_parent_4)[internal_node, right=0.25cm of internal_node_4] {4};
  \node(internal_parent_1)[gray_internal_node, left=0.25cm of internal_node_1] {1};
  \node(internal_parent_2)[gray_internal_node, right=0.25cm of internal_node_2] {2};
  \node(internal_parent_5)[internal_node, right=1.75cm of internal_node_5] {5};
  \node(internal_parent_6)[internal_node, left=0.25cm of internal_node_6] {6};
  \node(leaf_node_0)[gray_leaf_node] at ( 0.0,-4) {0};
  \node(leaf_node_1)[gray_leaf_node] at ( 1.5,-4) {1};
  \node(leaf_node_2)[gray_leaf_node] at ( 3.0,-4) {2};
  \node(leaf_node_3)[gray_leaf_node] at ( 4.5, -4) {3};
  \node(leaf_node_4)[gray_leaf_node] at ( 6.0, -4) {4};
  \node(leaf_node_5)[leaf_node] at ( 7.5, -4) {5};
  \node(leaf_node_6)[leaf_node] at ( 9.0, -4) {6};
  \node(leaf_node_7)[leaf_node] at (10.5, -4) {7};
  \draw [->] (internal_parent_0) edge   (internal_parent_3);
  \draw [->] (internal_parent_0) edge   (internal_parent_4);
  \draw [->] (internal_parent_3) edge   (internal_parent_1);
  \draw [->] (internal_parent_3) edge   (internal_parent_2);
  \draw [->] (internal_parent_4) edge   (leaf_node_4);
  \draw [->] (internal_parent_4) edge   (internal_parent_5);
  \draw [->] (internal_parent_1) edge   (leaf_node_0);
  \draw [->] (internal_parent_1) edge   (leaf_node_1);
  \draw [->] (internal_parent_2) edge   (leaf_node_2);
  \draw [->] (internal_parent_2) edge   (leaf_node_3);
  \draw [->] (internal_parent_5) edge   (internal_parent_6);
  \draw [->] (internal_parent_5) edge   (leaf_node_7);
  \draw [->] (internal_parent_6) edge   (leaf_node_5);
  \draw [->] (internal_parent_6) edge   (leaf_node_6);
\end{tikzpicture}
  \caption{An example of the tree traversal mask for a thread corresponding to
  a point with index 4.}\label{f:pair_traversal}
\end{figure}

We use an additional optimization in the main phase. In \Cref{a:fdbscan}, the
algorithm can be seen as operating on the edges of the adjacency graph. As the
results of $\union(x,y)$ and $\union(y,x)$ are identical from a cluster
membership perspective, it is sufficient to process each edge only once. To
facilitate this, we introduced a new hierarchy traversal algorithm. Given a
thread corresponding to a point with index $i$, a subtree corresponding to the
leaf nodes with indices less than $i$ is hidden from the thread. This way, the
thread avoids entering the subtrees with lower leaf indices, guaranteeing that
all the found neighbors would have indices $j > i$, thus guaranteeing that each
pair of neighboring points is processed exactly once. \Cref{f:pair_traversal}
demonstrates the tree mask for a thread corresponding to index 4. The thread
would stay in the right subtree of the root, skipping the left subtree
entirely. The advantages of such an approach include fewer memory accesses used
during the traversal, reduced number of distance computations, and reduced
number of \unionfind operations.

\subsection{FDBSCAN-DenseBox}\label{s:fdbscan_dense}

A given combination of \minpts{} and $\eps$ often results in the number of
neighbors within an $\eps$-neighborhood of a point significantly exceeding the
value of $\minpts$. In this case, many of the distance computations may be
avoided. In this Section, we propose an alternative approach to \fdbscan which
takes advantage of this fact.

Eliminating extra distance computations has been studied
in~\cite{welton2013,gowanlock2019a}. The methods operate by superimposing a
uniform Cartesian grid and processing cells with at least \minpts{} points more
efficiently. We integrated these ideas into a tree-based search index, which we
call \fdbscandense.

\begin{figure}
  \centering
    \includegraphics[width=0.2\textwidth]{figures/dense_cells.tex}\hfill
    \raisebox{0.4\height}{\includegraphics[width=0.26\textwidth]{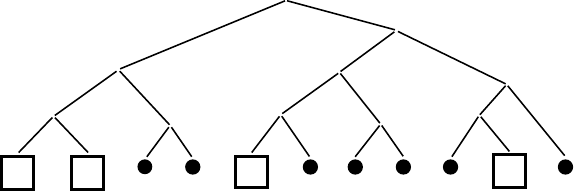}}
    \caption{Left: regular grid with grid size $\eps/\sqrt{d}$ superimposed over
    the dataset. The dense cells for $\minpts = 5$ are shown in red. Right: BVH
	constructed from a mixed set of objects.}\label{f:dense_cells}
\end{figure}

The procedure starts with computing the bounds of the data set and imposing a
regular grid over the computational domain. The grid cell length is set to be
$\eps/\sqrt{d}$, with $d$ being the data dimension. This choice guarantees that
the diameter of each cell does not exceed $\eps$. Next, we calculate a cell
index for all points in the dataset, and determine the number of points in each
cell. The cells with at least \minpts{} are called \emph{dense}.
\Cref{f:dense_cells} demonstrates a grid superimposed over a set of points,
with dense cells for $\minpts = 5$ marked in red.
It is clear that all the points in the dense cells are core points, and belong
to the same cluster. Thus, the distance calculations among the points in the
same dense cell can be eliminated.

The number of dense cells and the number of points inside them depend heavily
on the dataset data distribution and the parameters $\eps$ and $\minpts$. If the
value of $\eps$ is small compared to the domain size, the number of grid cells
in each dimension may be in thousands or more, resulting in billions of grid
cells. The data is then spread across a relatively small population of non-empty
cells. Searching for nearby cells in this situation becomes non-trivial. While it
is possible to do a series of binary searches over a list of cells to
produce a list of neighboring non-empty cells, in this work we use an
alternative approach.

To accommodate dense boxes, we modify the BVH construction algorithm of \fdbscan.
In \fdbscandense, the hierarchy is constructed out of a mix of points outside of
dense cells and the boxes of the dense cells. This is possible to do as the BVH
only requires bounding volumes for a set of objects.
Thus, such mixing does not impose any additional constraints. The use of this
approach with other trees, such as \kdtree, would pose more challenges.

Given the knowledge that all points in dense cells are core points, only the
points outside of dense cells have to be examined to identify the remaining
core points in the preprocessing phase. For every such point, the algorithm
finds all nearby objects within distance $\eps$ using the BVH. If the found
object is an isolated point, the neighbor count is incremented by one. If it is
a box (corresponding to a dense cell), a linear search over all points in that
cell is performed, incrementing the count each time a point is within distance
$\eps$. Similar to \fdbscan, the neighbors are only counted until reaching the
\minpts{} threshold, after which the procedure terminates.

At the beginning of the main phase, the \union operation is executed for all
points within the same dense cell. Then, the neighborhood search is performed
for all points in the dataset. During the search, once an object within
distance $\eps$ is found for an individual point, one of two cases may happen.
In the first case, the found object is a dense box. In this case, it is
sufficient to determine whether a single point of that dense box is within
distance $\eps$. A thread checks the distances to all points in that dense cell
linearly, until either a point within $\eps$ is found, in which case $\union()$
is called, or all points are exhausted. In the second case, the found object is
another point (outside of any dense cell). As the newly found point is within
$\eps$, the usual resolution depending on the core status of both points is
executed.

One drawback of \fdbscandense, compared to \fdbscan, is its use of arithmetic
operations (e.g., summation) when dealing with the cell computations. These
calculations may suffer from a loss of precision in the situations where the
value of $\eps$ is tiny compared to the coordinates of the data points,
potentially resulting in the erroneous results. This should be detected and
guarded against in an implementation. Alternatively, this could be addressed by
using a higher precision floating point numbers, or through the hashing
techniques. \fdbscan, on the other hand, only uses \textsc{min} and
\textsc{max} operations on the user data and does not have this limitation.

\subsection{\unionfind}

We chose the algorithm proposed in~\cite{jaiganesh2018} as our
\unionfind approach, being synchronization-free on
GPUs. Like most efficient implementations, it uses pointer jumping, a technique
to shorten paths of the trees (associated with disjoint sets) during the \find
operation. Specifically, the work uses ``intermediate pointer jumping'', which
compresses the path of all elements encountered on a way to the tree root by
making every element skip over the next element,
halving the path length in each traversal. Because the path compression does
not guarantee that all paths are fully compressed at the end of the main phase
(i.e., that the label of each point in the same cluster is identical
at the end of the main phase), an extra  finalization phase is introduced to
make each point directly to the representative.

\section{Experimental results }\label{s:results}

\ifjournal
\begin{table}[!t]
\else
\begin{table*}[!t]
\fi
\centering
\footnotesize
\caption{Datasets and the default parameters}\label{t:datasets}
  \begin{tabular}{lp{6pt}p{8pt}clccc}
\toprule
    \multirow{2}{*}{Name} & \multirow{2}{*}{$d$} & \multirow{2}{*}{$n$}       & \multirow{2}{*}{Source}            & \multirow{2}{*}{Description} & \multicolumn{3}{c}{Default parameters} \\
    \cmidrule(lr){6-8} &     &           &                   &             & $\eps$                & \minpts & Samples \\
\midrule
  2D-NGSIM            & 2   & $\sim$12M & ~\cite{ngsim}     & GPS loc     & 1.0                   & 10      & 100K \\
  2D-Porto            & 2   & $\sim$81M & ~\cite{portotaxi} & GPS loc     & 0.005                 & 10      & 100K \\
  2D-SS-simden        & 2   & 10M       & ~\cite{gan2017}   & Generated   & 1000                  & 10      & 100K \\
  2D-SS-varden        & 2   & 10M       & ~\cite{gan2017}   & Generated   & 1000                  & 10      & 100K \\
  3D-Hacc             & 3   & $\sim$37M & ~\cite{hacc}      & Cosmology   & 0.042                 & 10      & 1M \\
  3D-SS-simden        & 3   & 10M       & ~\cite{gan2017}   & Generated   & 1000                  & 10      & 1M \\
  3D-SS-varden        & 3   & 10M       & ~\cite{gan2017}   & Generated   & 1000                  & 10      & 1M \\
  5D-SS-simden        & 5   & 10M       & ~\cite{gan2017}   & Generated   & 1000                  & 10      & 1M \\
  5D-SS-varden        & 5   & 10M       & ~\cite{gan2017}   & Generated   & 1000                  & 10      & 1M \\
  7D-SS-simden        & 7   & 10M       & ~\cite{gan2017}   & Generated   & 1000                  & 10      & 1M \\
  7D-SS-varden        & 7   & 10M       & ~\cite{gan2017}   & Generated   & 1000                  & 10      & 1M \\
  7D-Household        & 7   & $\sim$2M  & ~\cite{household} & Power       & 2.0                   & 10      & 1M \\
\bottomrule
\end{tabular}
\ifjournal
\end{table}
\else
\end{table*}
\fi

In our implementation, we used ArborX~\cite{arborx2020}, an open-source library
for the tree-based implementations using Kokkos library~\cite{kokkos2022} for
a device-independent programming model.
Kokkos offers parallel execution patterns (parallel loops,
reductions, scans) to abstract from a specific hardware. Kokkos also provides
abstractions for execution and memory resources. The Kokkos
library\footnote{\url{https://github.com/kokkos/kokkos}} provides C++
abstractions and supports hardware through backends, including Nvidia GPUs
(Cuda), AMD GPUs (HIP), and serial hosts (Serial).

\unless\ifblind
The implemented algorithms are available in the main ArborX
repository\footnote{\url{https://github.com/arborx/ArborX}}.
\fi

The ArborX library provides several features suitable for our implementation.
It allows for an early traversal termination, which is used in the
preprocessing phases of both \fdbscan and \fdbscandense. The callback
functionality of the library allows execution of a user-provided code on a
positive match, which is used both in preprocessing for the neighbor count and
in the main phase for the \unionfind kernels.

\noindent\textbf{Testing environment.}
The numerical studies presented in the paper were performed using \amdcpu (64
cores\footnote{Run as 56 cores, with 8 cores dedicated to OS processes}),
\nvidiagpu (40GB) and a single GCD (Graphics Compute Die) of
\amdgpu\footnote{Currently, HIP (Heterogeneous-computing Interface for
Portability) -- the programming interface provided by AMD -- only allows the
use of each GCD as an independent GPU.}. The chips are based on TSMC's N7+, N7
and N6 technology, respectively, and can be considered to belong to the same
generation.

We used Clang 14.0.0 compiler for \amdcpu, NVCC 11.5 for \nvidiagpu, and ROCm 5.4.3 for \amdgpu.

\noindent\textbf{Datasets.}
As mentioned in~\Cref{s:introduction}, in this work we focus on the
low-dimensional data. For our experiments, we used a combination of artificial
and real-world datasets listed in~\Cref{t:datasets} to comprehensively evaluate
our algorithm and meet our study goals. The GPS locations (\dataset{2D-NGSIM} and
\dataset{2D-Porto}), cosmology (\dataset{3D-HACC}) and electric power consumption
(\dataset{7D-Household}) datasets replicate real-world conditions. The
datasets generated with~\cite{gan2017} allow us to explore more structure and
dimensionalities. \dataset{SS-simden} and \dataset{SS-varden} refer to the
datasets with similar-density and variable-density clusters, respectively.

\subsection{Parallel algorithms comparison}\label{s:2D}

In this Section, we compare the performance of \fdbscan and \fdbscandense
algorithms with the several other implementations: \gdbscan~\cite{andrade2013}
(only available for 2D datasets), \pdsdbscan~\cite{patwary2015} and
\tepp~\cite{wang2020}. We did not include the results for
CUDA-DClust~\cite{bohm2009} as it was many orders of magnitude slower.
Unfortunately, we were also not able to compare to the recent
CUDA-DClust+~\cite{gowanlock2021} code\footnote{\url{https://github.com/l3lackcurtains/fast-cuda-gpu-dbscan}}, as it
consistently produced wrong results and did not match the performance reported
in~\cite{gowanlock2021}; the problem seems to be related to
\texttt{thrust::equal\_range} routines and is being investigated by the original
authors at the time of this publication.

We study the behavior of the algorithms for each dataset varying one of the
three parameters, $\eps$, \minpts, and the number of drawn random samples,
while keeping the other two fixed at the default values shown
in~\Cref{t:datasets}. The default number of samples for the 2D datasets was
chosen to be lower to accomodate \gdbscan's memory consumption.

In this Section, the \gdbscan, \fdbscan and \fdbscandense experiments were
performed on \nvidiagpu.

\subsubsection*{Impact of $\eps$}

\begin{figure*}[!t]
    \centering
    \includegraphics[width=0.6\textwidth]{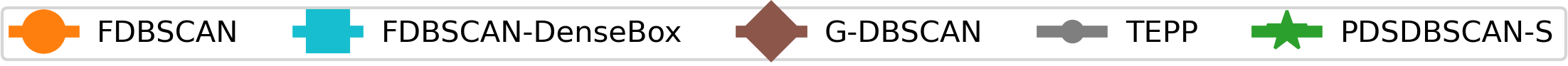} \\ [-2ex]

    \subfloat[2D-NGSIM]{\includegraphics[width=0.24\textwidth]{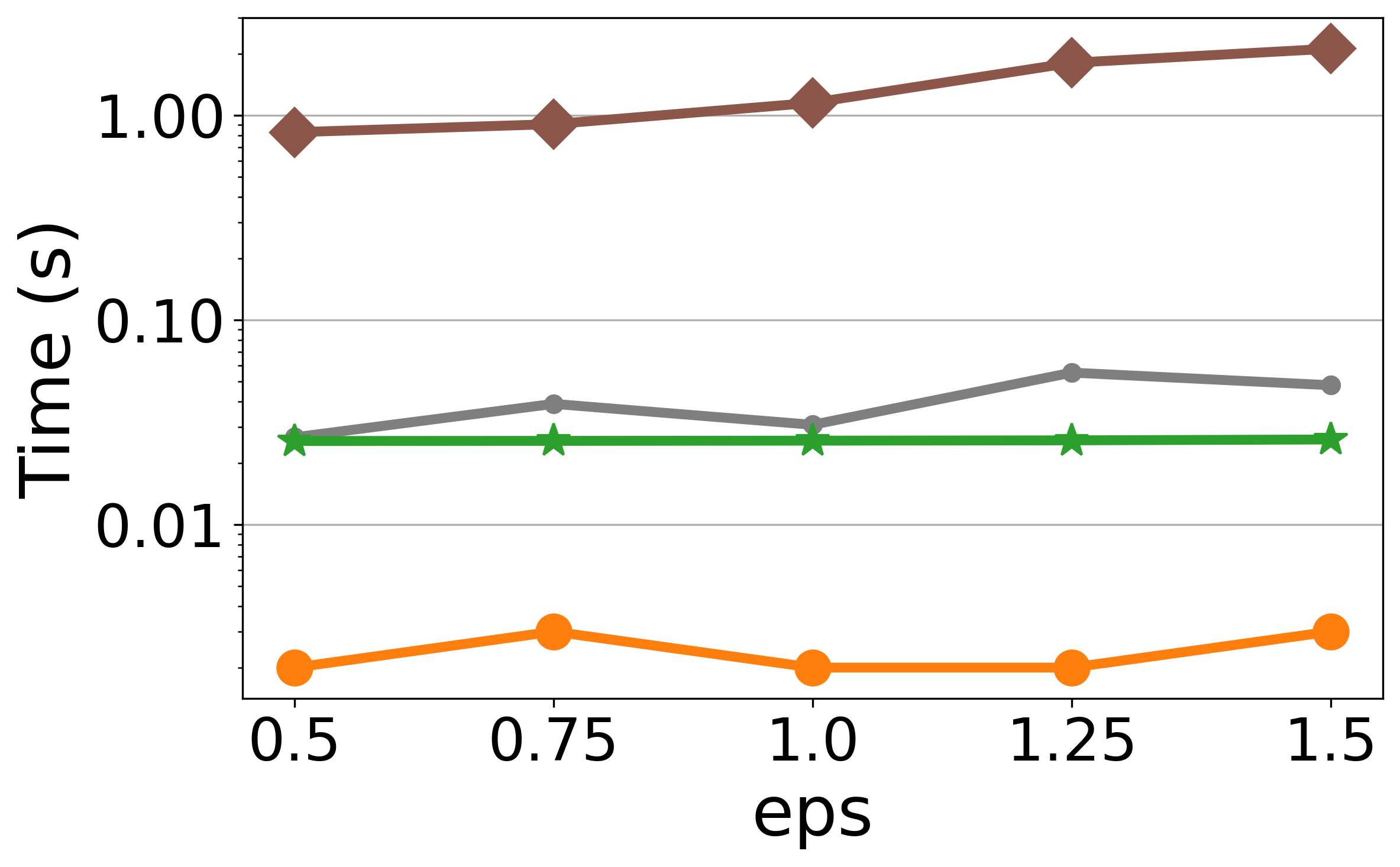}}
    \subfloat[2D-Porto]{\includegraphics[width=0.24\textwidth]{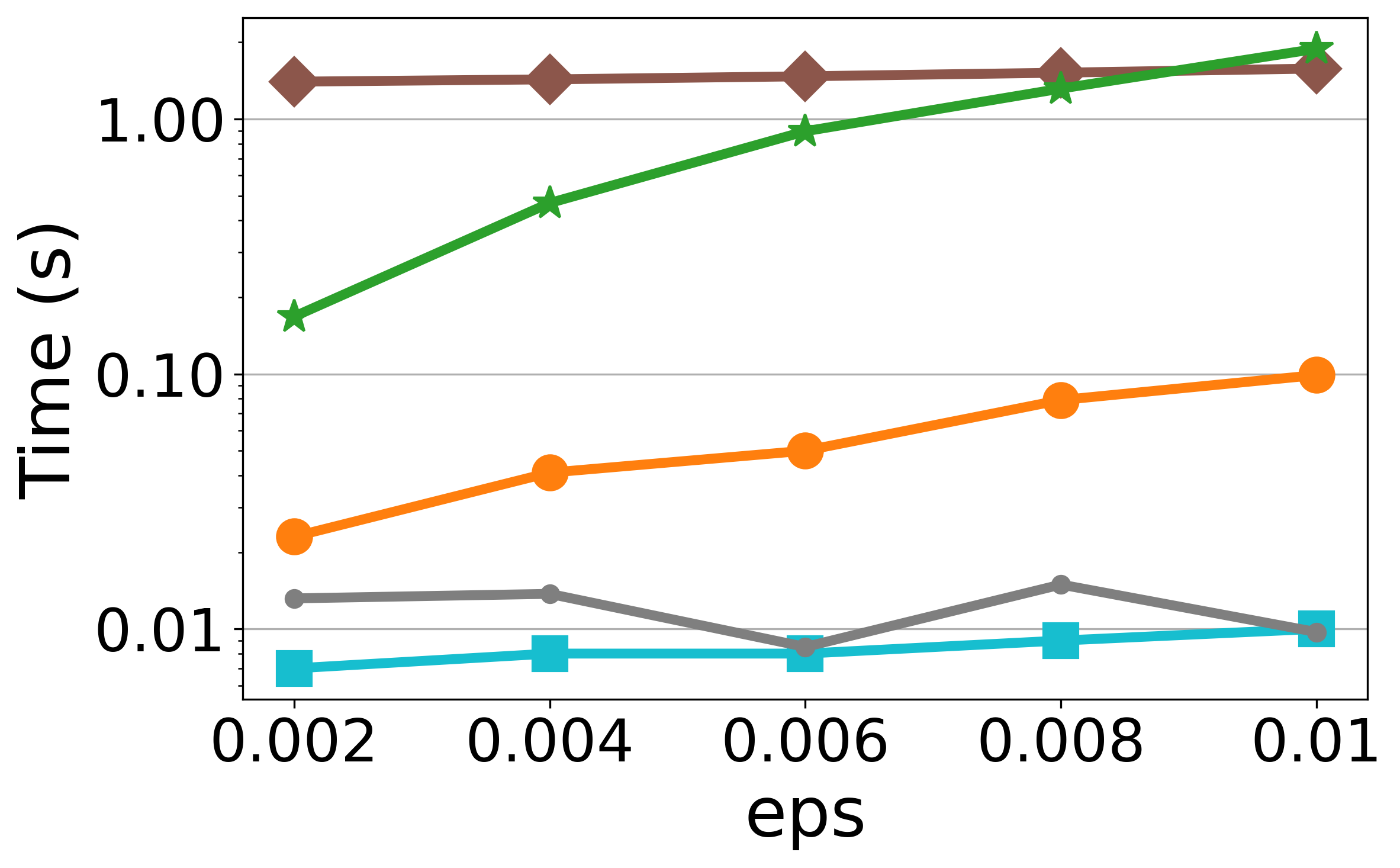}}
    \subfloat[2D-SS-simden]{\includegraphics[width=0.24\textwidth]{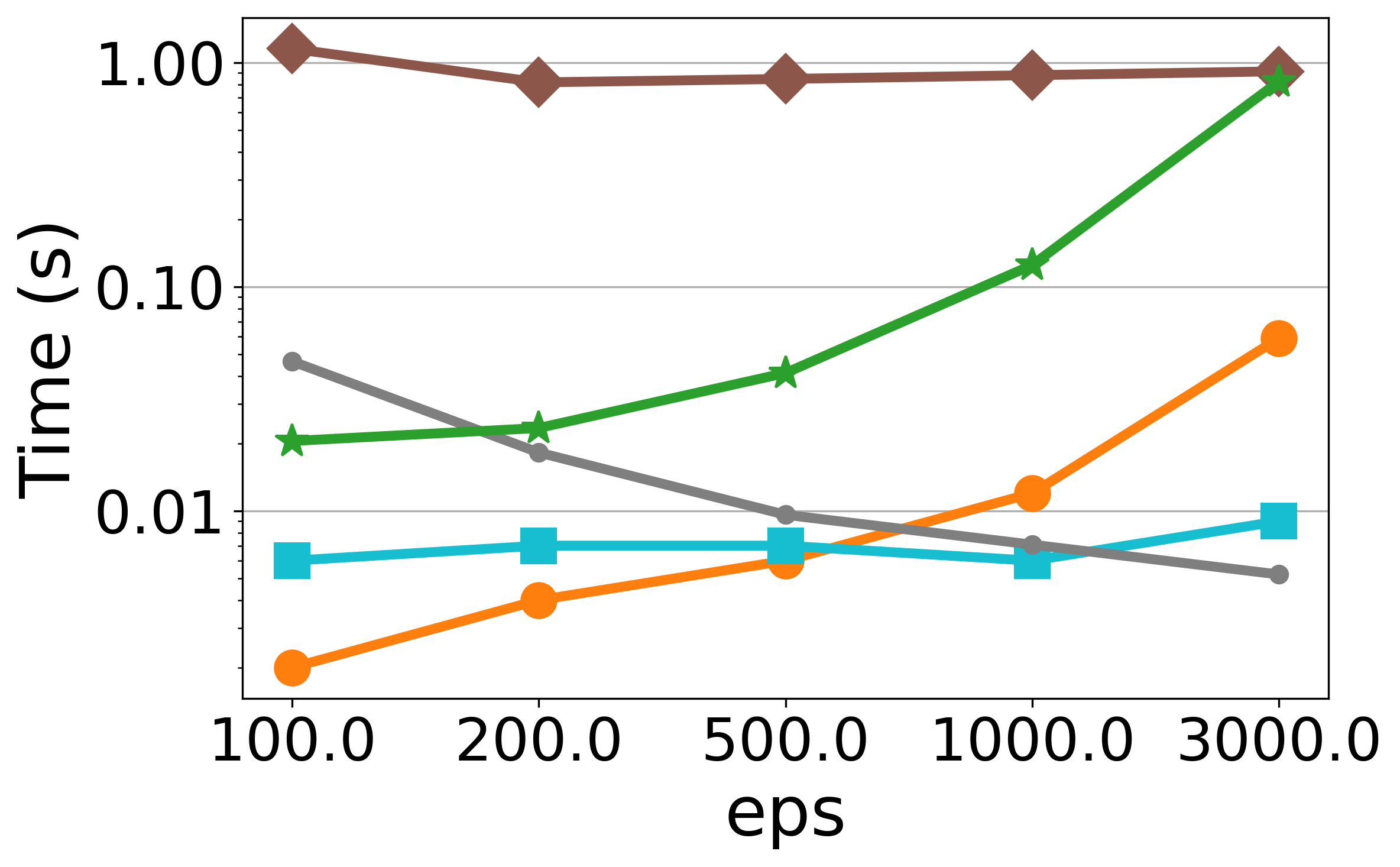}}
    \subfloat[2D-SS-varden]{\includegraphics[width=0.24\textwidth]{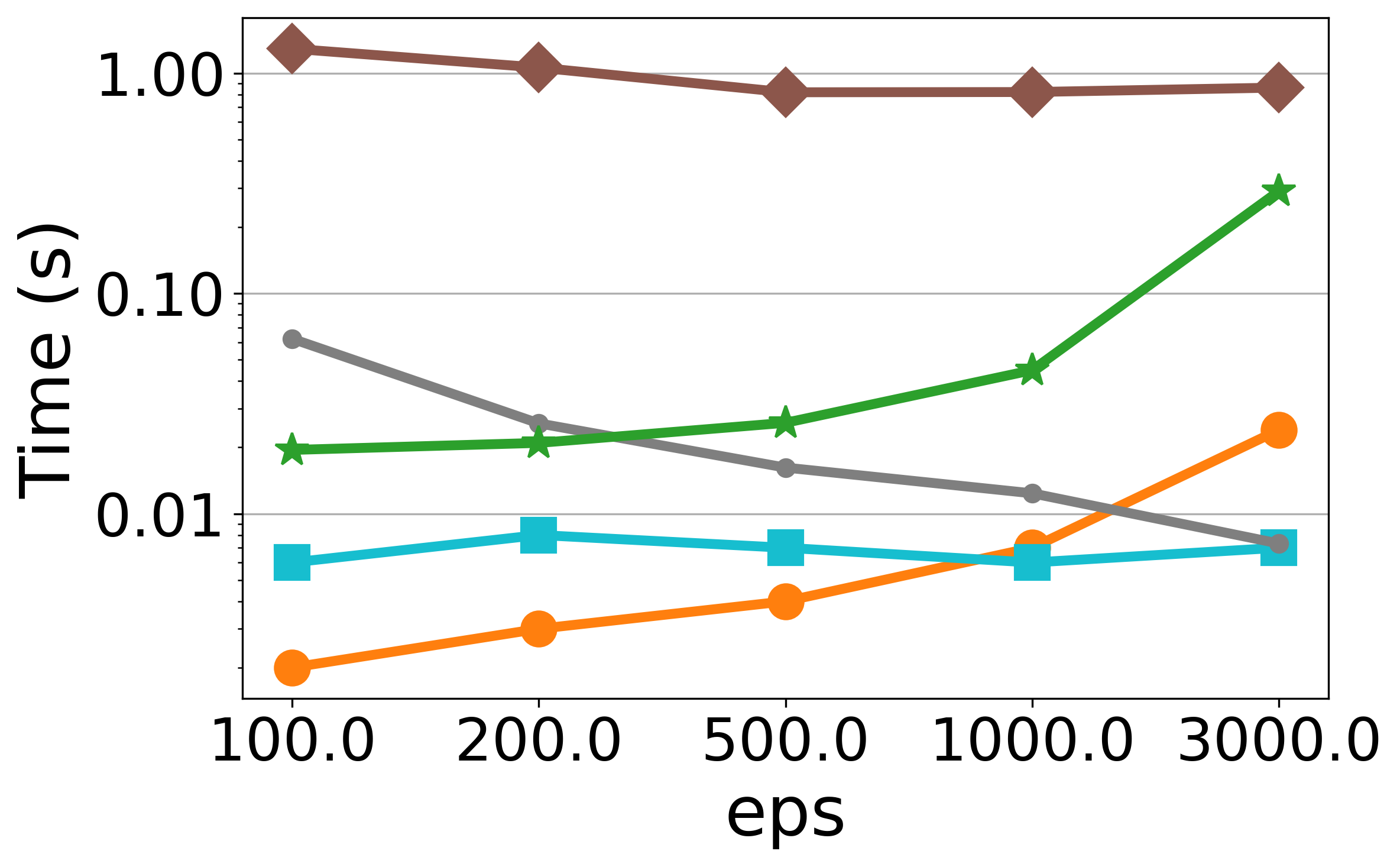}} \\[-2ex]

    \subfloat[3D-Hacc]{\includegraphics[width=0.24\textwidth]{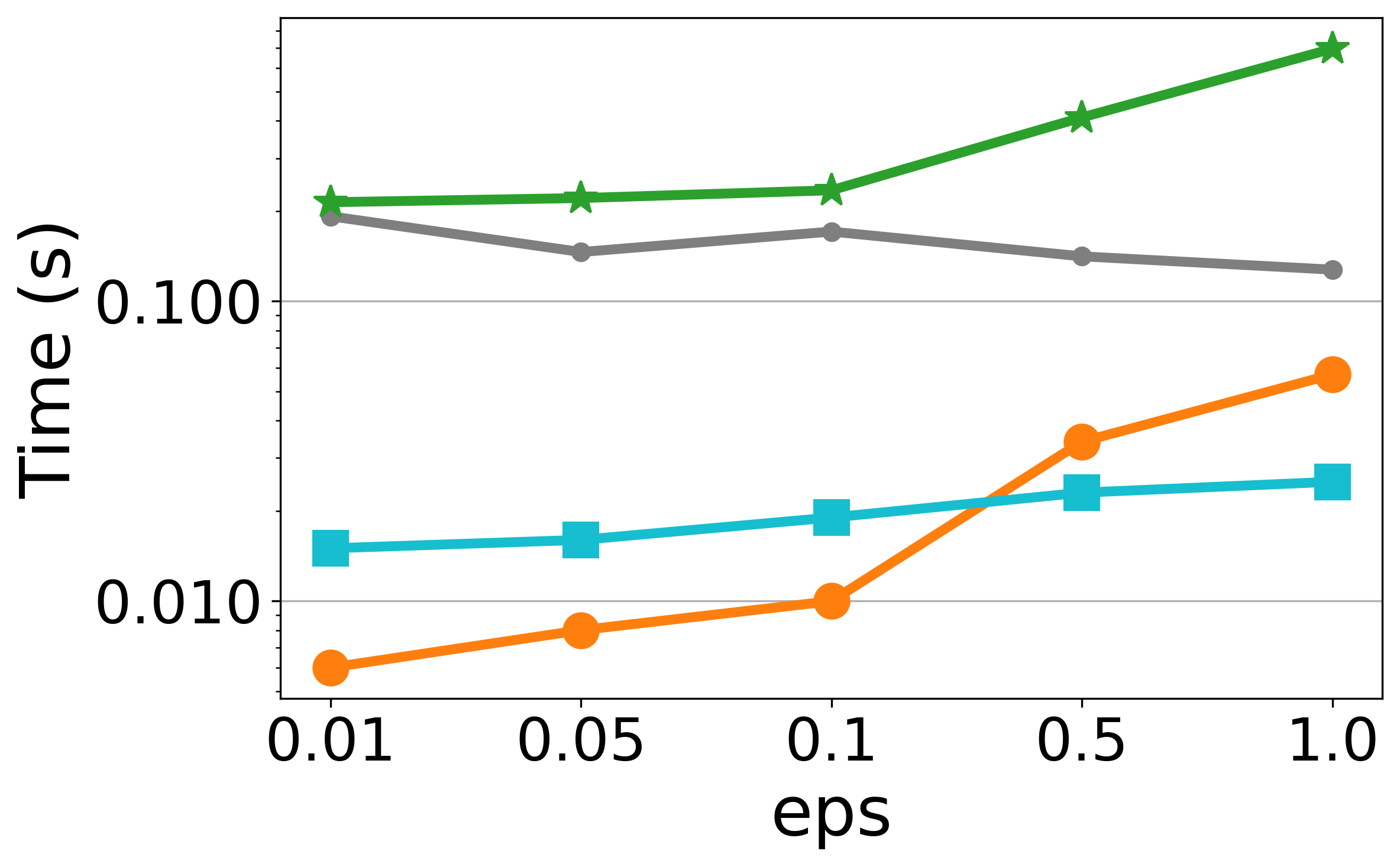}}
    \subfloat[3D-SS-varden]{\includegraphics[width=0.24\textwidth]{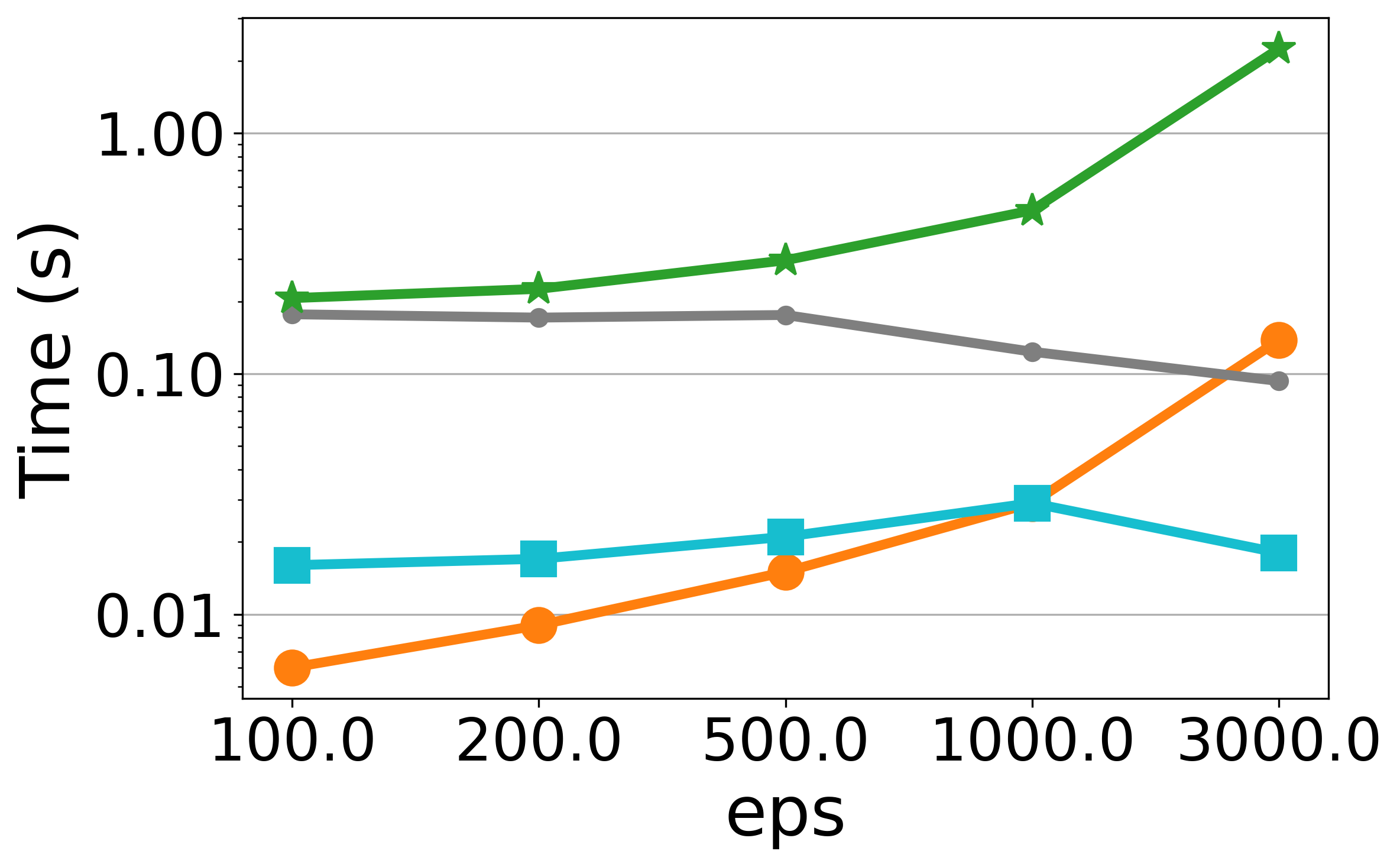}}
    \subfloat[3D-SS-varden]{\includegraphics[width=0.24\textwidth]{figures/3D-ss-var_eps.png}}
    \subfloat[5D-SS-simden]{\includegraphics[width=0.24\textwidth]{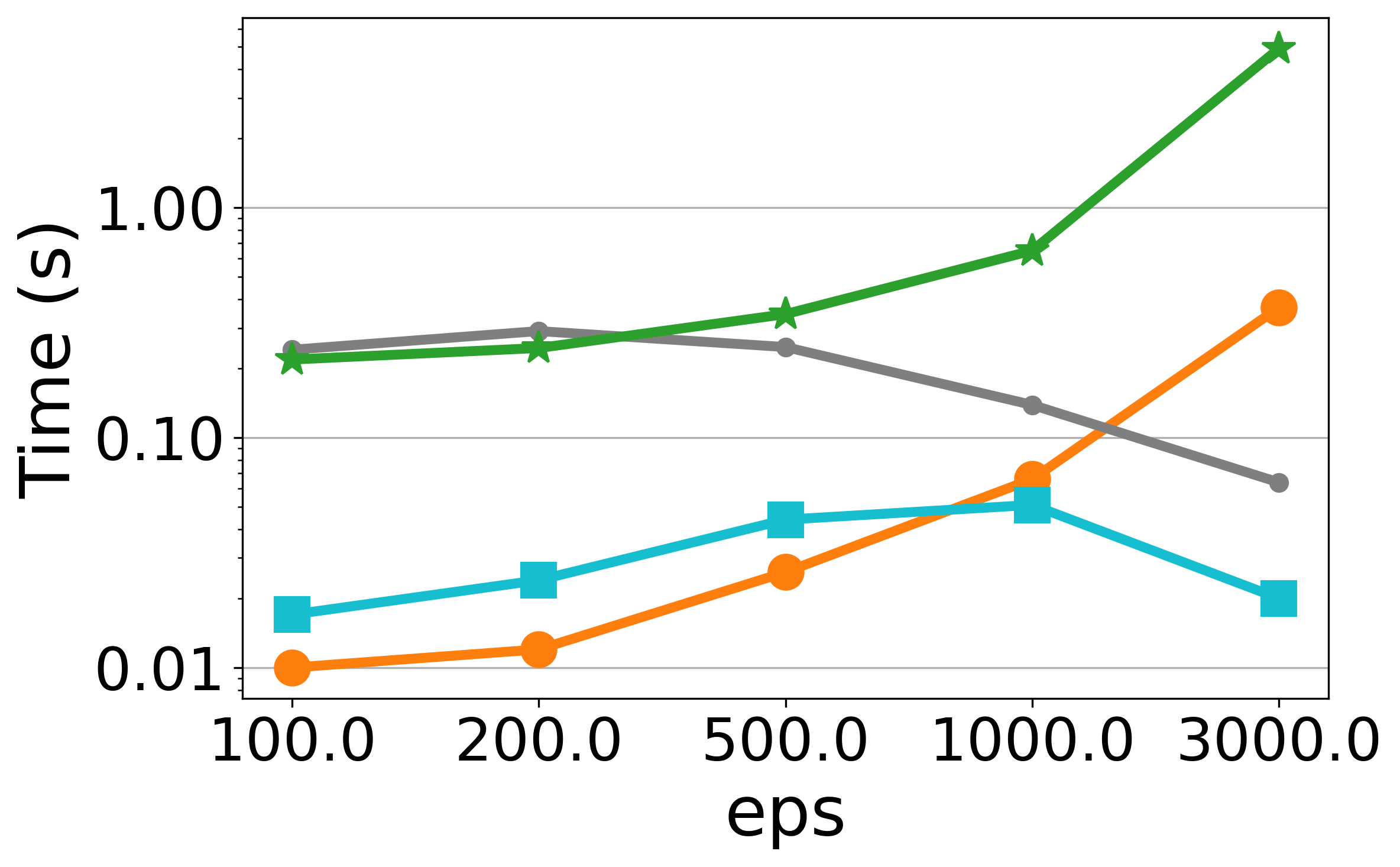}} \\[-2ex]

    \subfloat[5D-SS-varden]{\includegraphics[width=0.24\textwidth]{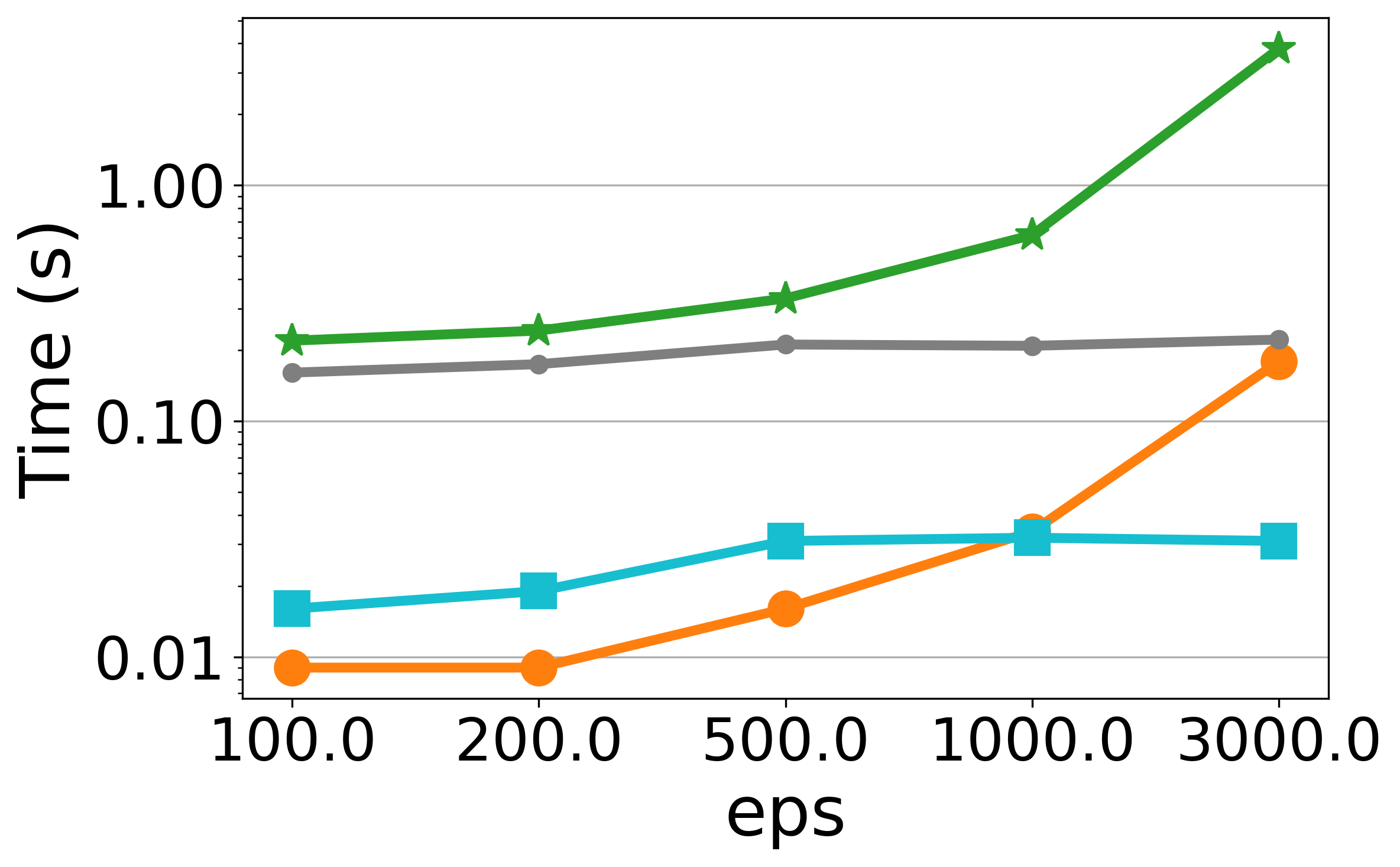}}
    \subfloat[7D-SS-simden]{\includegraphics[width=0.24\textwidth]{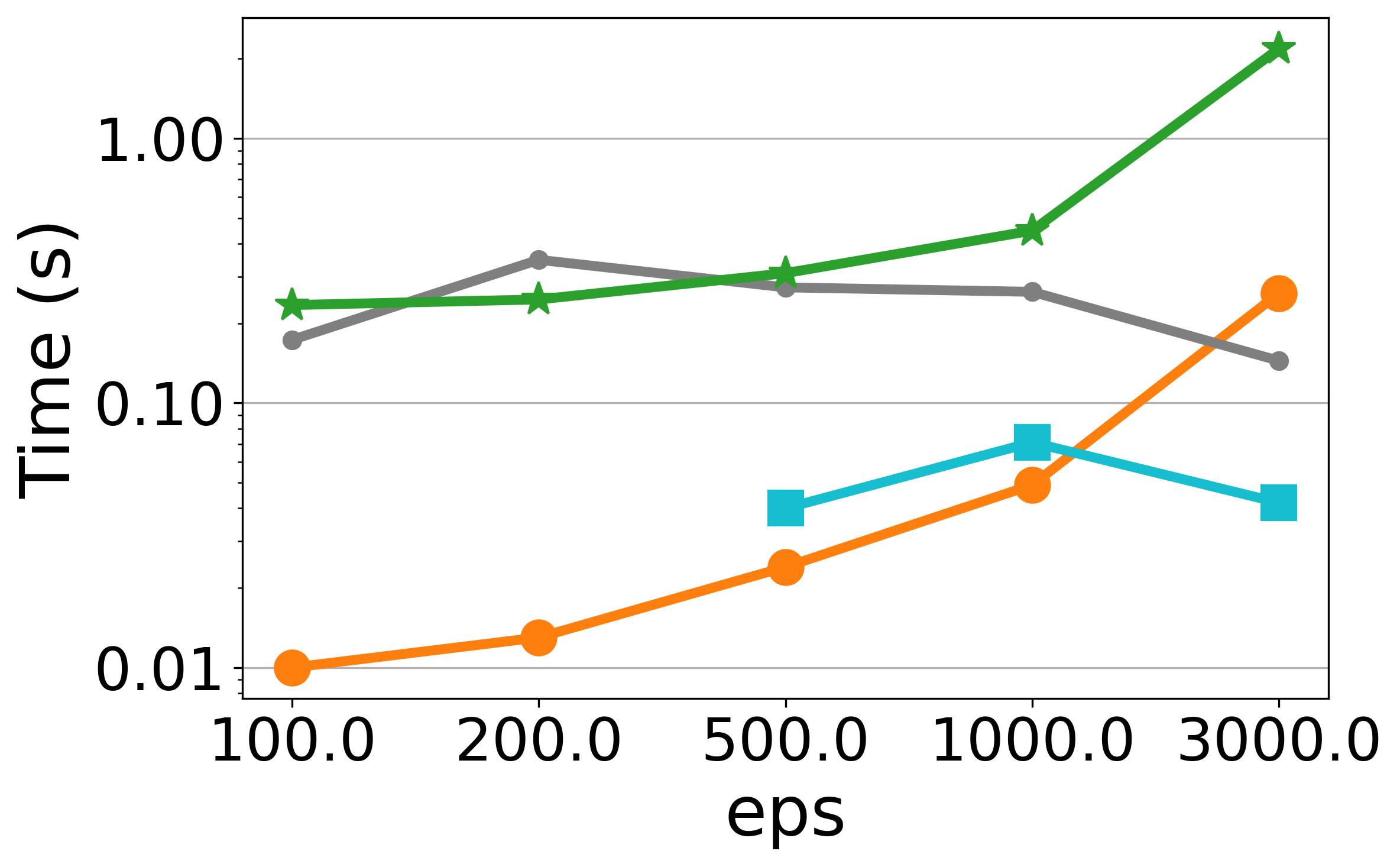}}
    \subfloat[7D-SS-varden]{\includegraphics[width=0.24\textwidth]{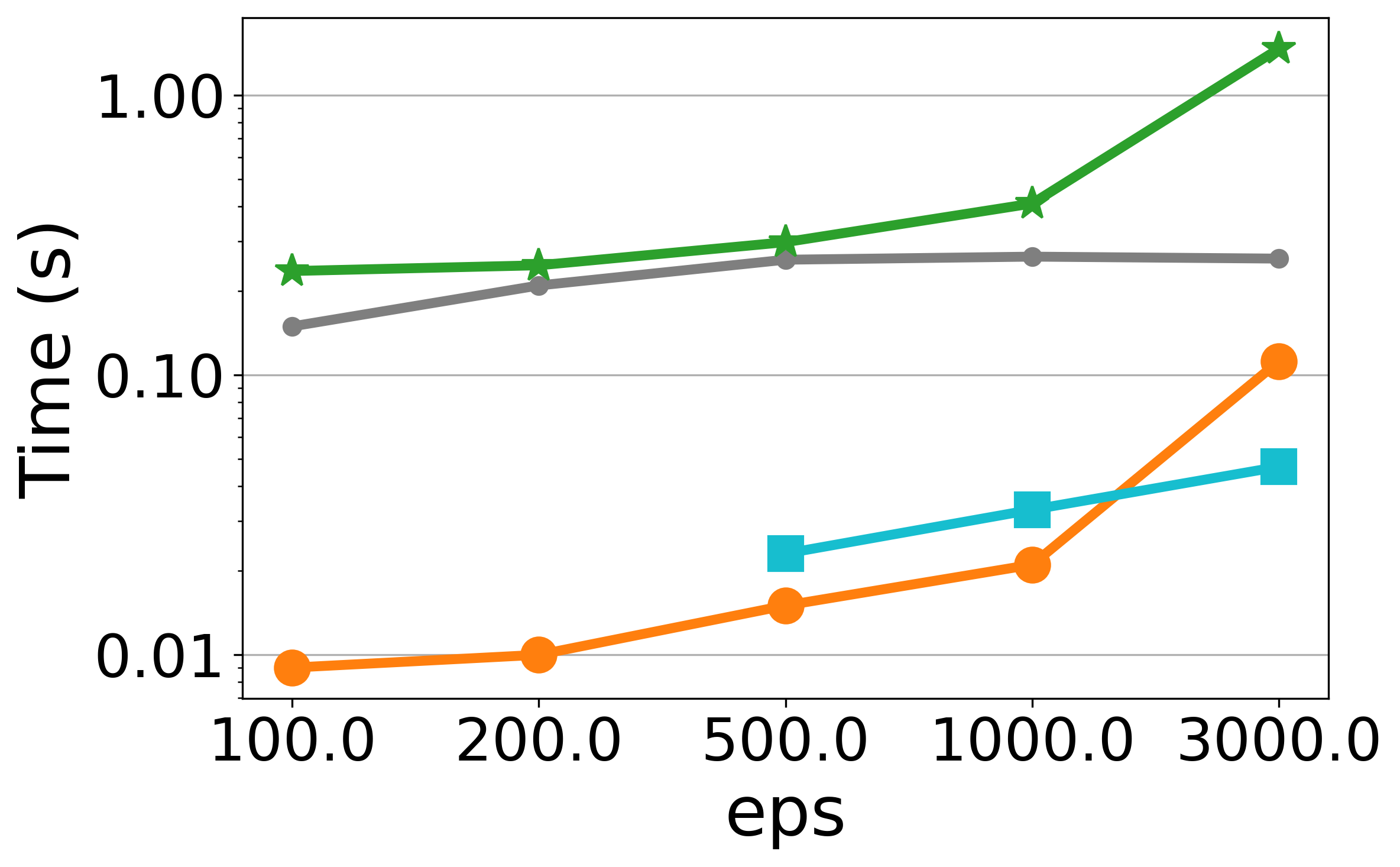}}
    \subfloat[7D-Household]{\includegraphics[width=0.24\textwidth]{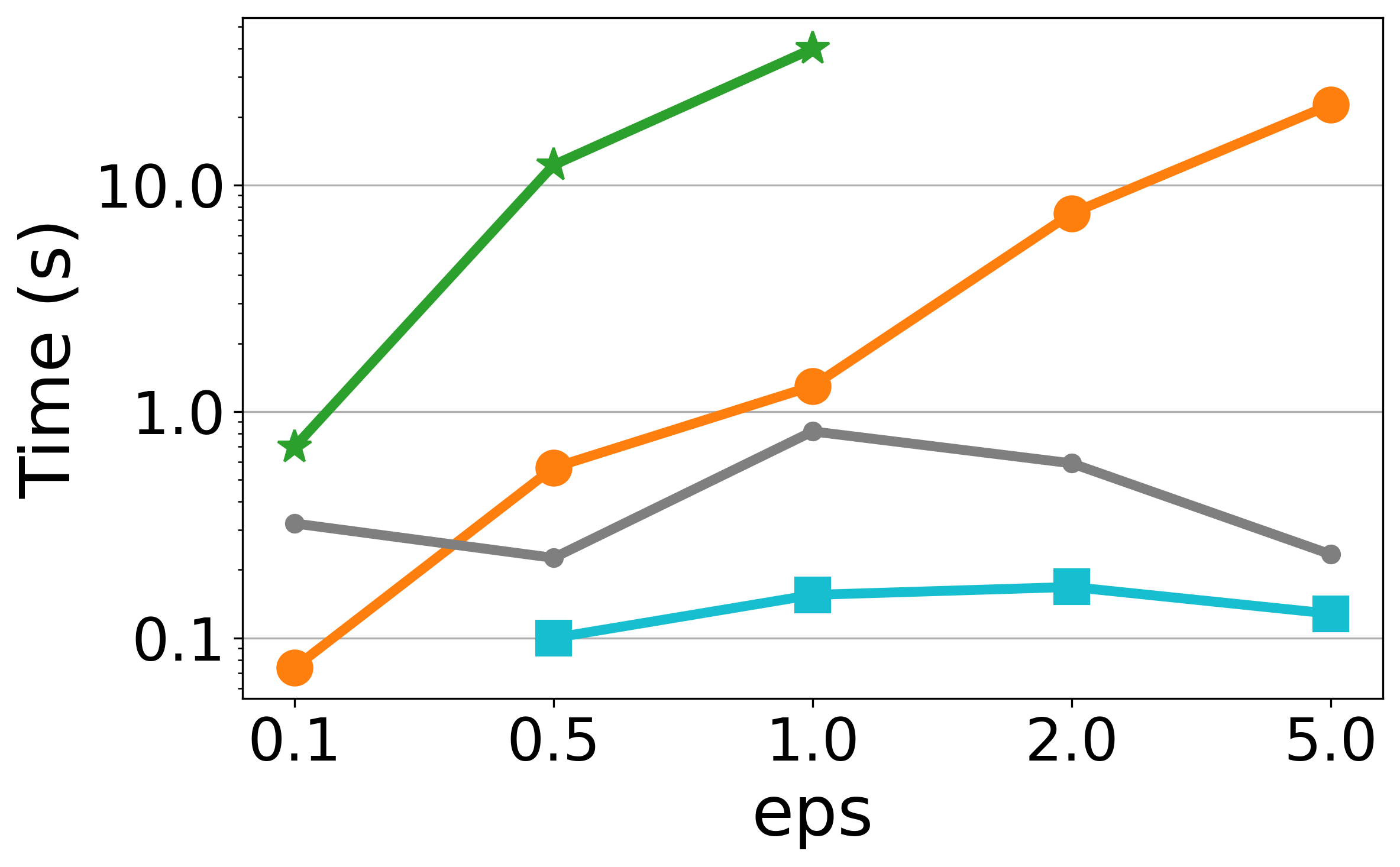}}

    \caption{Impact of the $\eps{}$ parameter on the execution time.}\label{f:sweep_eps}
\end{figure*}

\Cref{f:sweep_eps} demonstrates the impact of the parameter $\eps$ on the
execution times while keeping $\minpts{}$ and problem size fixed at the default
values. Increasing $\eps$ increases the size of each neighborhood $N_\eps(x)$,
thus increasing the cluster sizes. The range of $\eps$ for each problem was
chosen in such a way that the number of clusters qualitatively changes from
many small clusters to a few large ones.

We first observe that for the 2D cases where \gdbscan was able to run, it is an
obvious outlier in terms of performance. Generally, \pdsdbscan is the second
slowest, running significantly slower than \tepp and \fdbscan, particularly for
larger values of $\eps$. \tepp is competitive with \fdbscan in some situations,
particularly for large values of $\eps$ and \dataset{2D-Porto}, where the
densities of the data points are high and \fdbscan performs a lot of unnecessary
computations. However, \fdbscandense outperforms \tepp in almost all situations,
except for the \dataset{2D-Porto} and the largest values of $\eps$ for
\dataset{2D-SS-simden}, \dataset{2D-SS-varden}. \fdbscan outperforms
\fdbscandense for lower values of $\eps$ in most situations, which corresponds
to situations with lower density values, and thus few (if any) dense cells. The
rule of thumb is to use \fdbscan for very low sparsity situations, and
\fdbscandense otherwise.

It is important to note the few missing data points in the plots. First, we see
that \fdbscandense is completely missing in \dataset{2D-NGSIM}, and in several
lower $\eps$ values for \dataset{7D-SS-simden} and \dataset{7D-SS-varden}. As
we mentioned at the end of \Cref{s:fdbscan_dense}, a combination of the domain
size and the values of $\eps$ may lead to \fdbscandense losing precision and
potentially leading to the wrong results. This is exactly what is happening here,
and our implementation of \fdbscandense aborted the computation. We also
observe missing data for \pdsdbscan for \dataset{7D-Household}, where it ran
out of memory.

Another interesting observation is the expected dependence of \fdbscan on
the $\eps$ parameter: larger values of $\eps$ result in the longer runtimes, as
it increases the size of $N_\eps(x)$ neighborhoods, and \fdbscan has no
mechanisms to avoid additional computations. On the other hand, the time for
\fdbscandense is relatively stable for the full range of $\eps$.

\subsubsection*{Impact of \minpts}

\begin{figure*}[!t]
    \centering
    \includegraphics[width=0.5\textwidth]{figures/sweep_legend.png} \\ [-2ex]
    \subfloat[2D-NGSIM]{\includegraphics[width=0.24\textwidth]{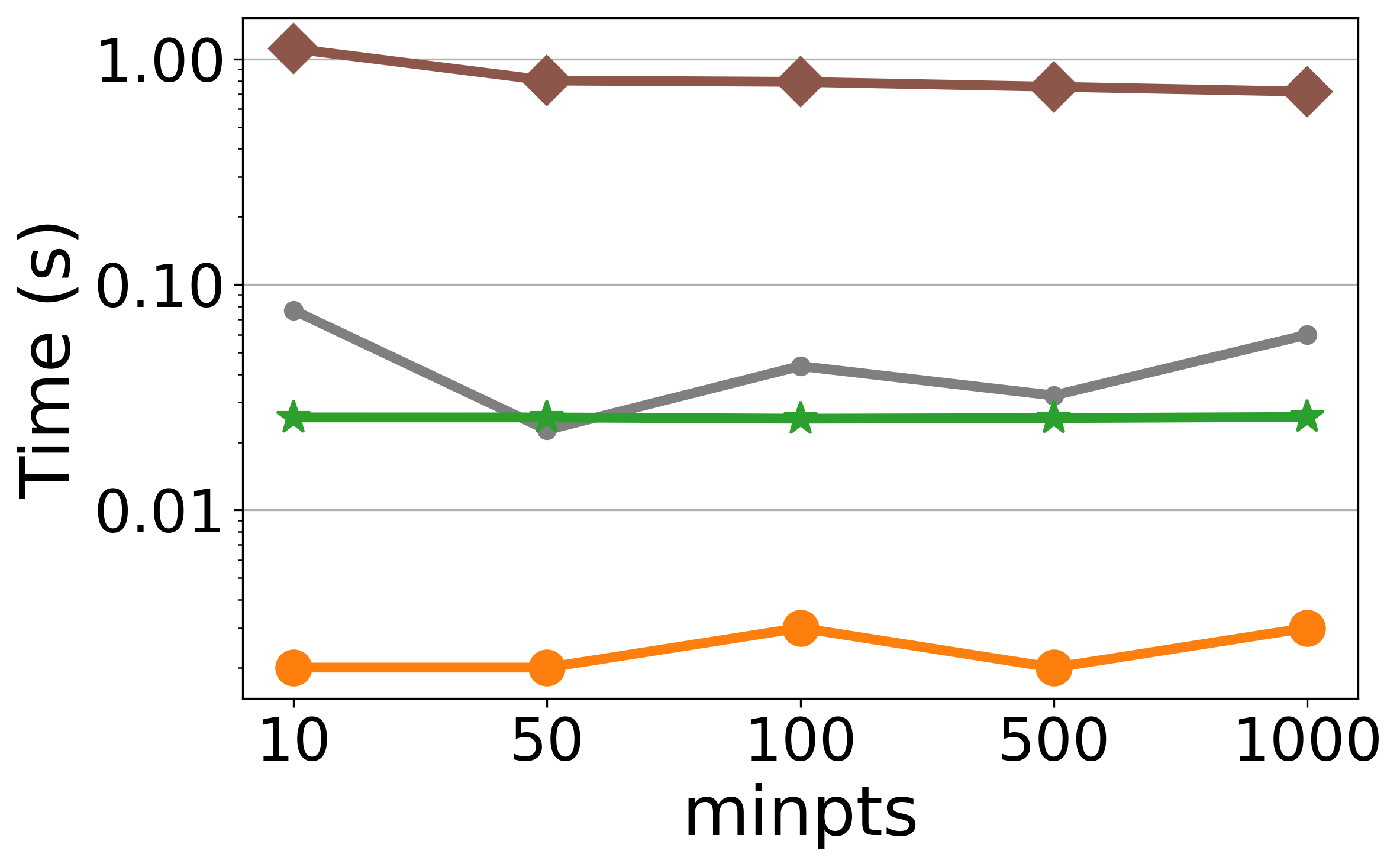}}
    \subfloat[2D-Porto]{\includegraphics[width=0.24\textwidth]{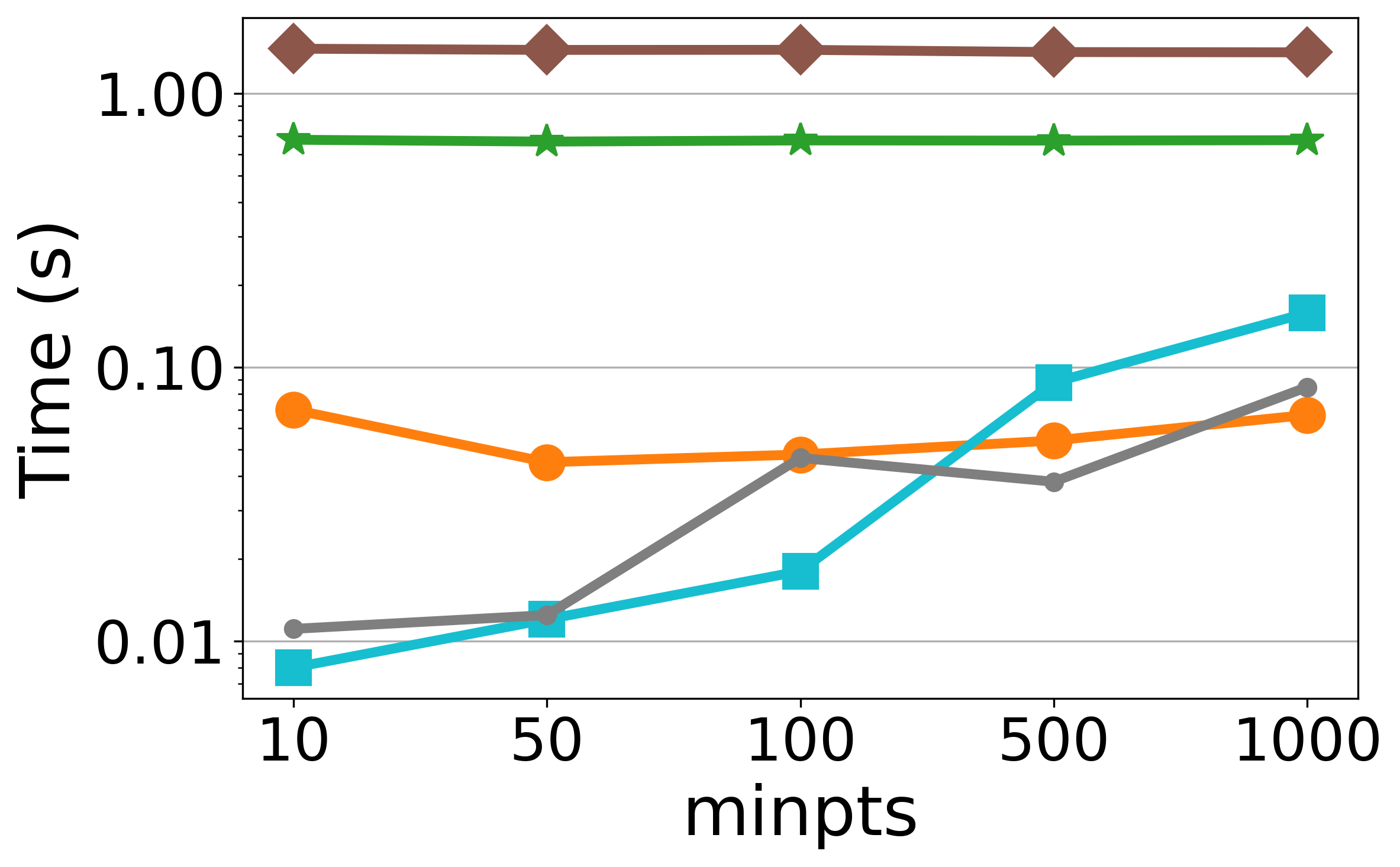}}
    \subfloat[2D-SS-simden]{\includegraphics[width=0.24\textwidth]{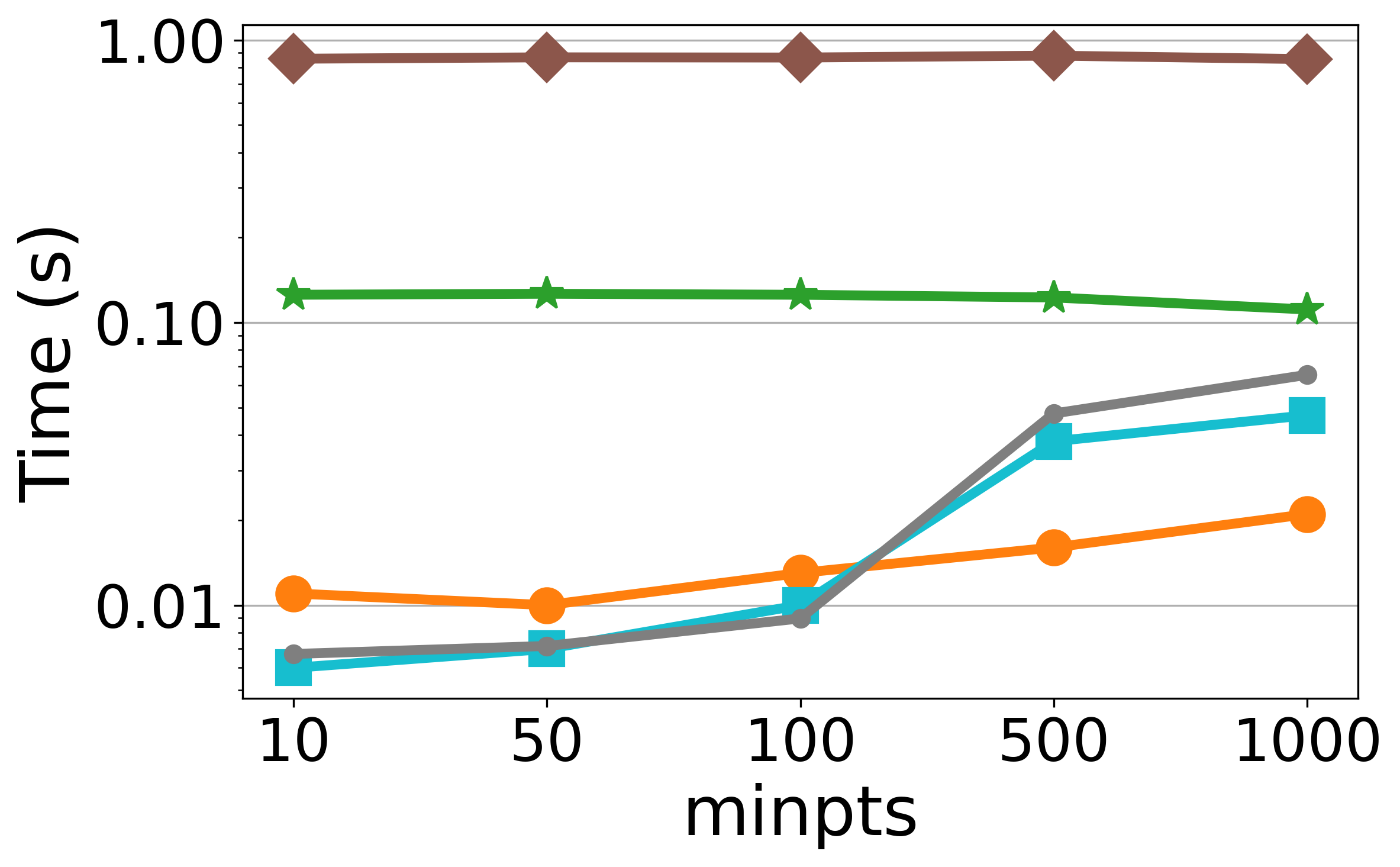}}
    \subfloat[2D-SS-varden]{\includegraphics[width=0.24\textwidth]{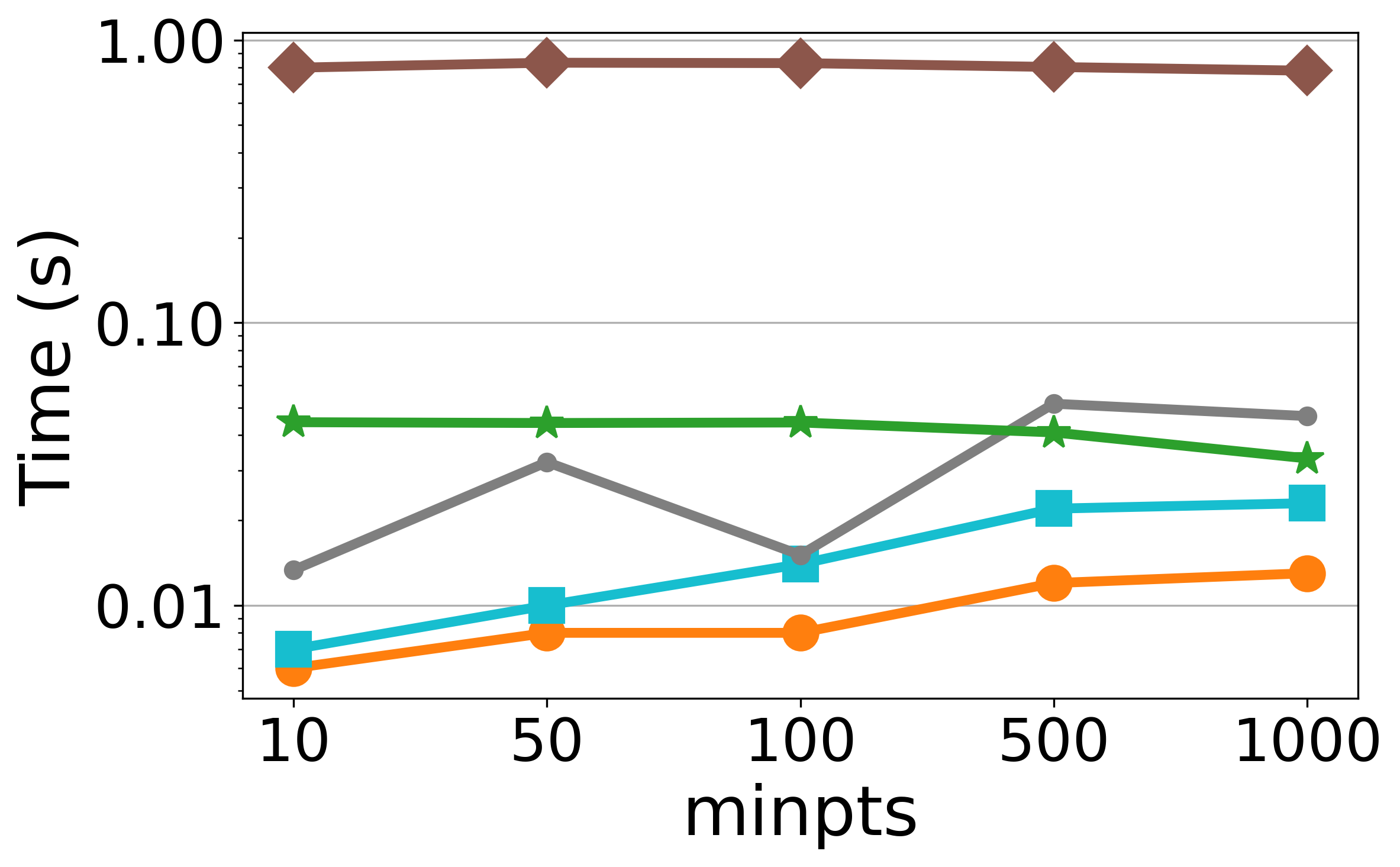}} \\[-2ex]

    \subfloat[3D-Hacc]{\includegraphics[width=0.24\textwidth]{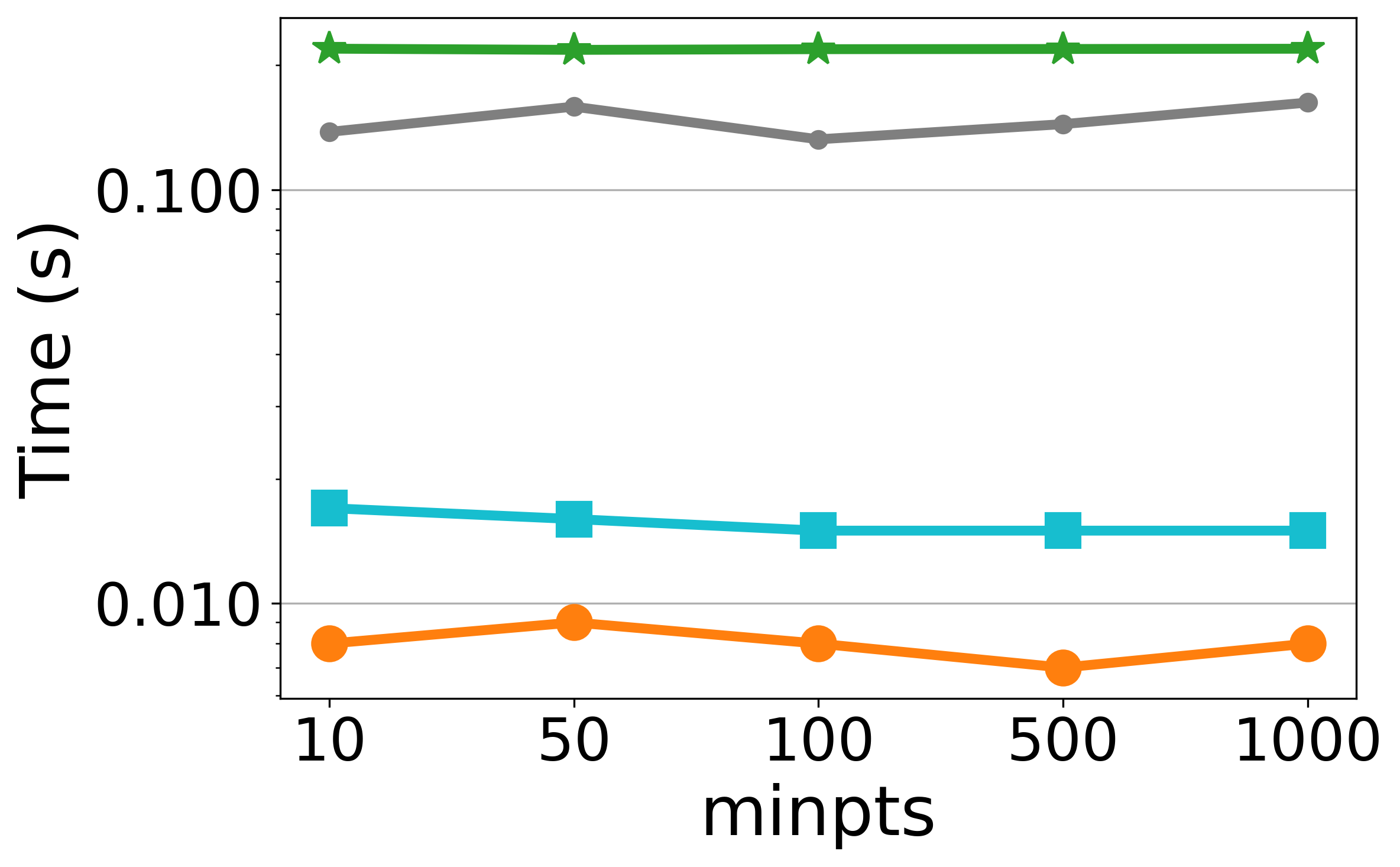}}
    \subfloat[3D-SS-varden]{\includegraphics[width=0.24\textwidth]{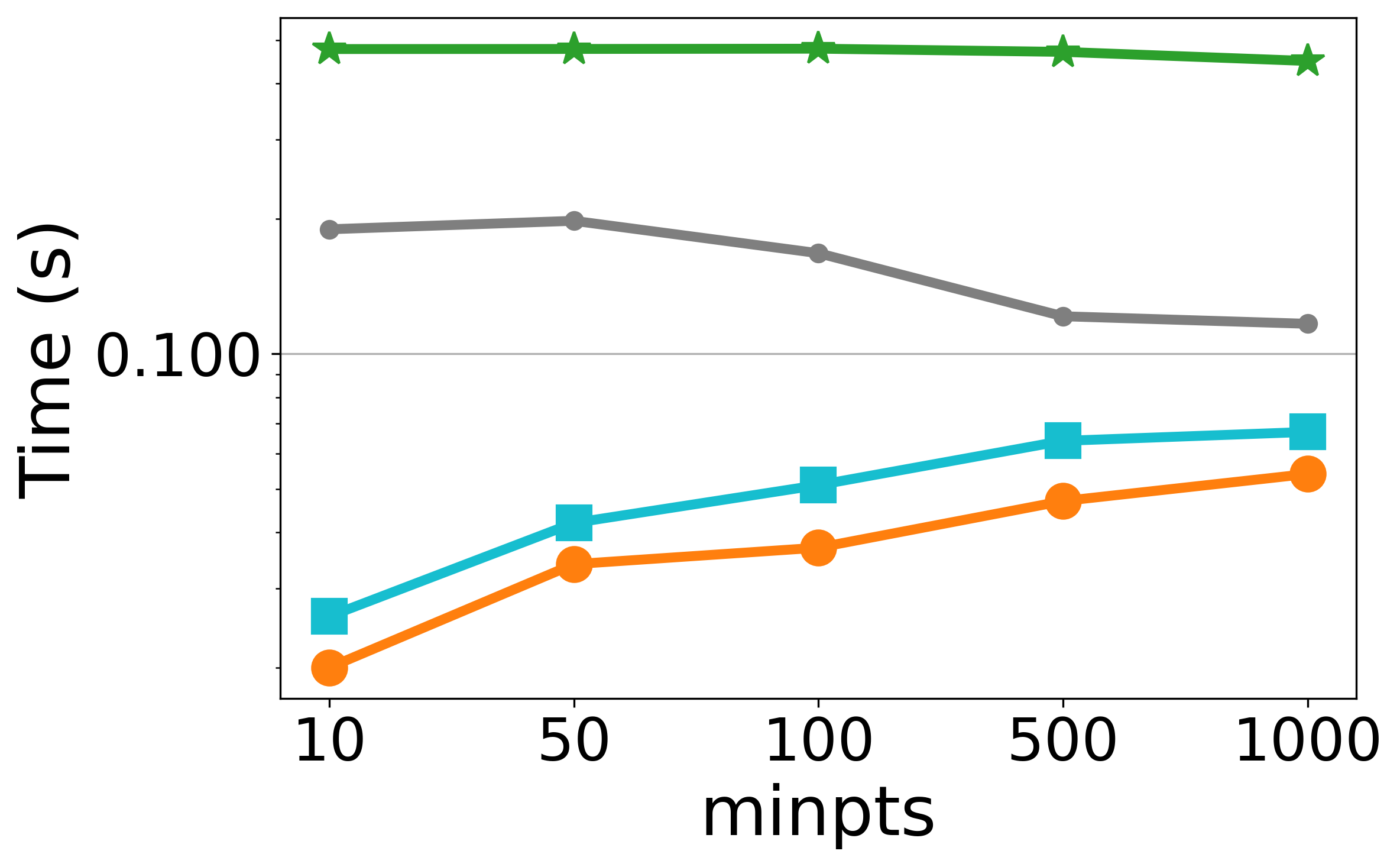}}
    \subfloat[3D-SS-varden]{\includegraphics[width=0.24\textwidth]{figures/3D-ss-var_minpts.png}}
    \subfloat[5D-SS-simden]{\includegraphics[width=0.24\textwidth]{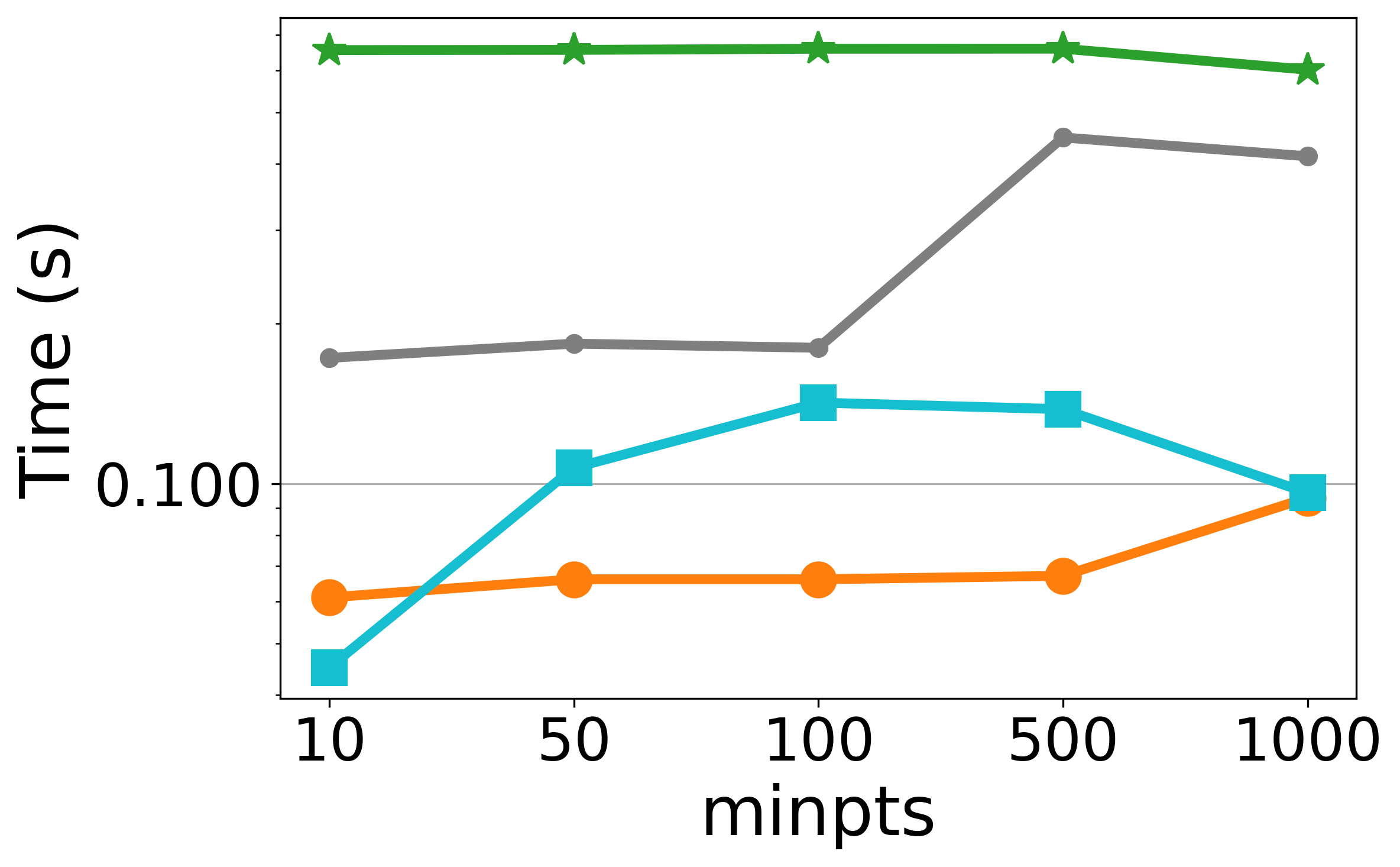}} \\[-2ex]

    \subfloat[5D-SS-varden]{\includegraphics[width=0.24\textwidth]{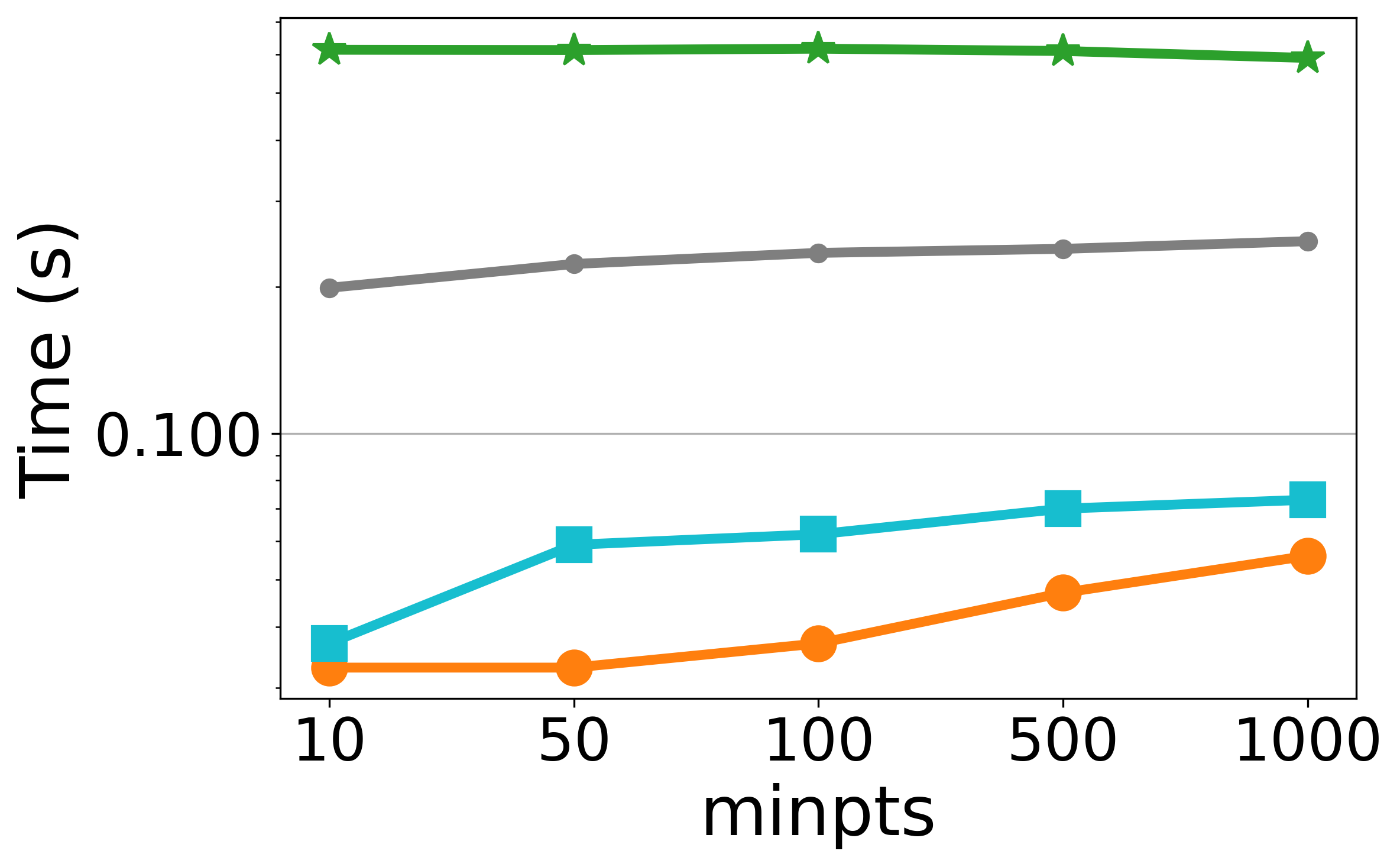}}
    \subfloat[7D-SS-simden]{\includegraphics[width=0.24\textwidth]{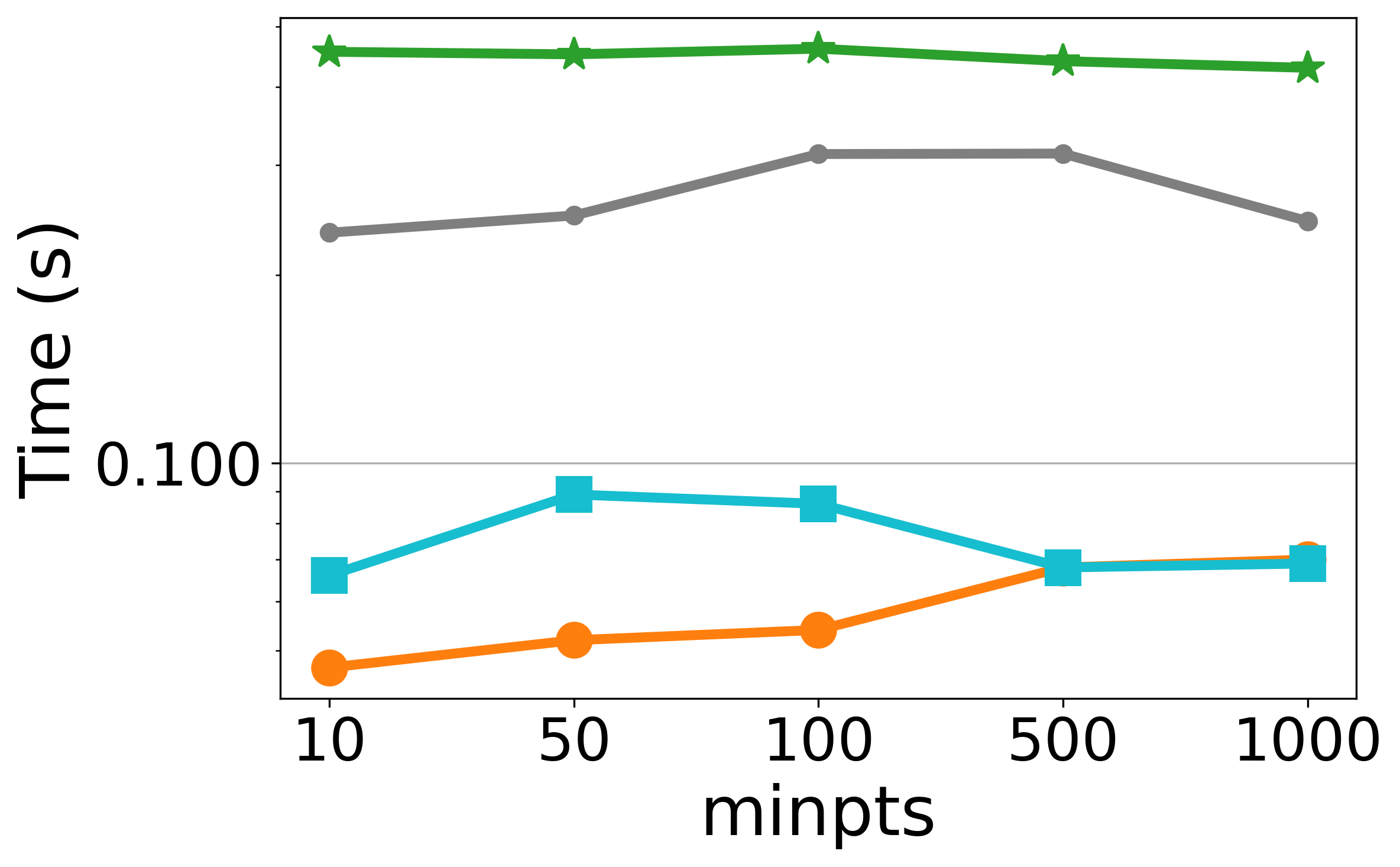}}
    \subfloat[7D-SS-varden]{\includegraphics[width=0.24\textwidth]{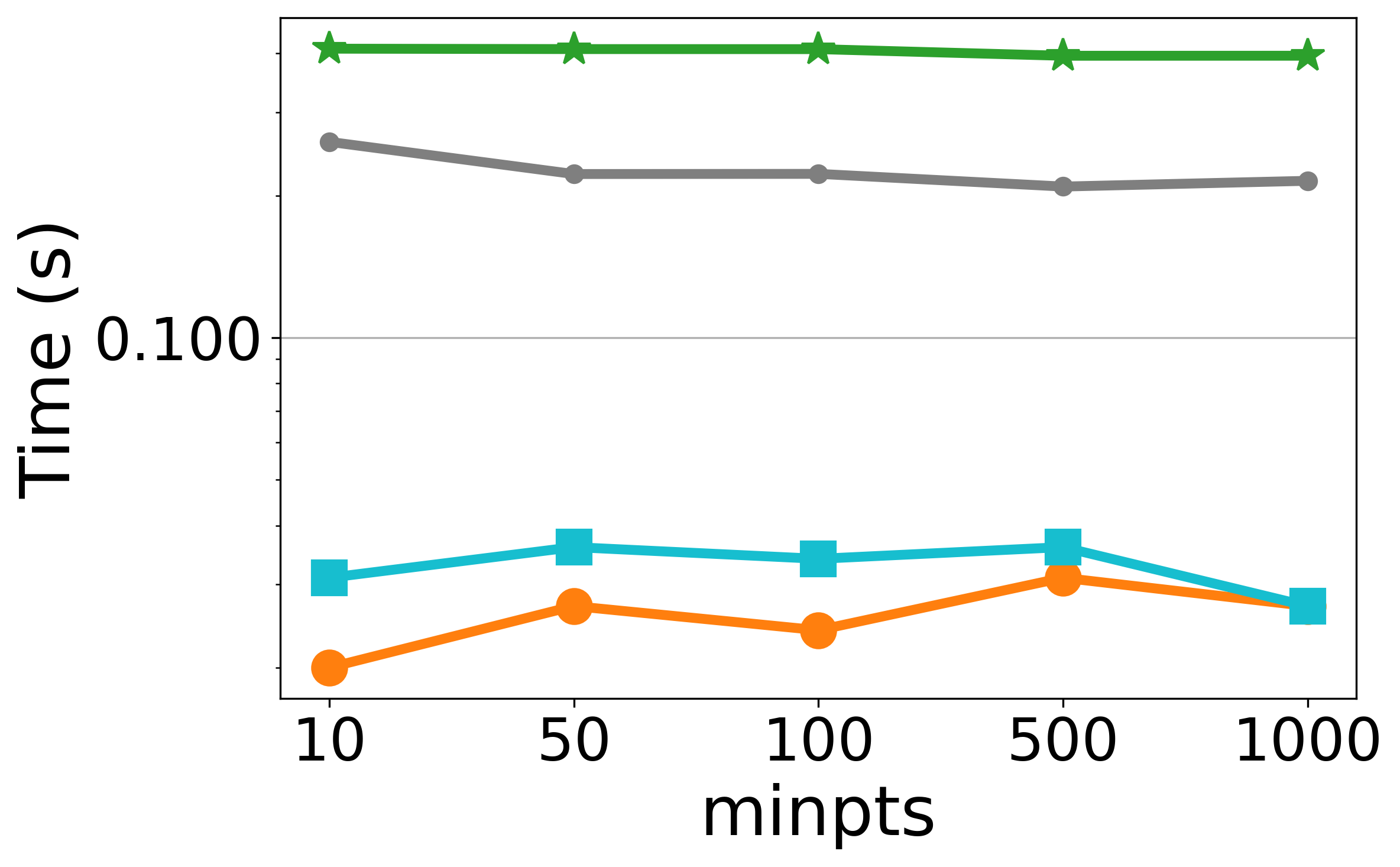}}
    \subfloat[7D-Household]{\includegraphics[width=0.24\textwidth]{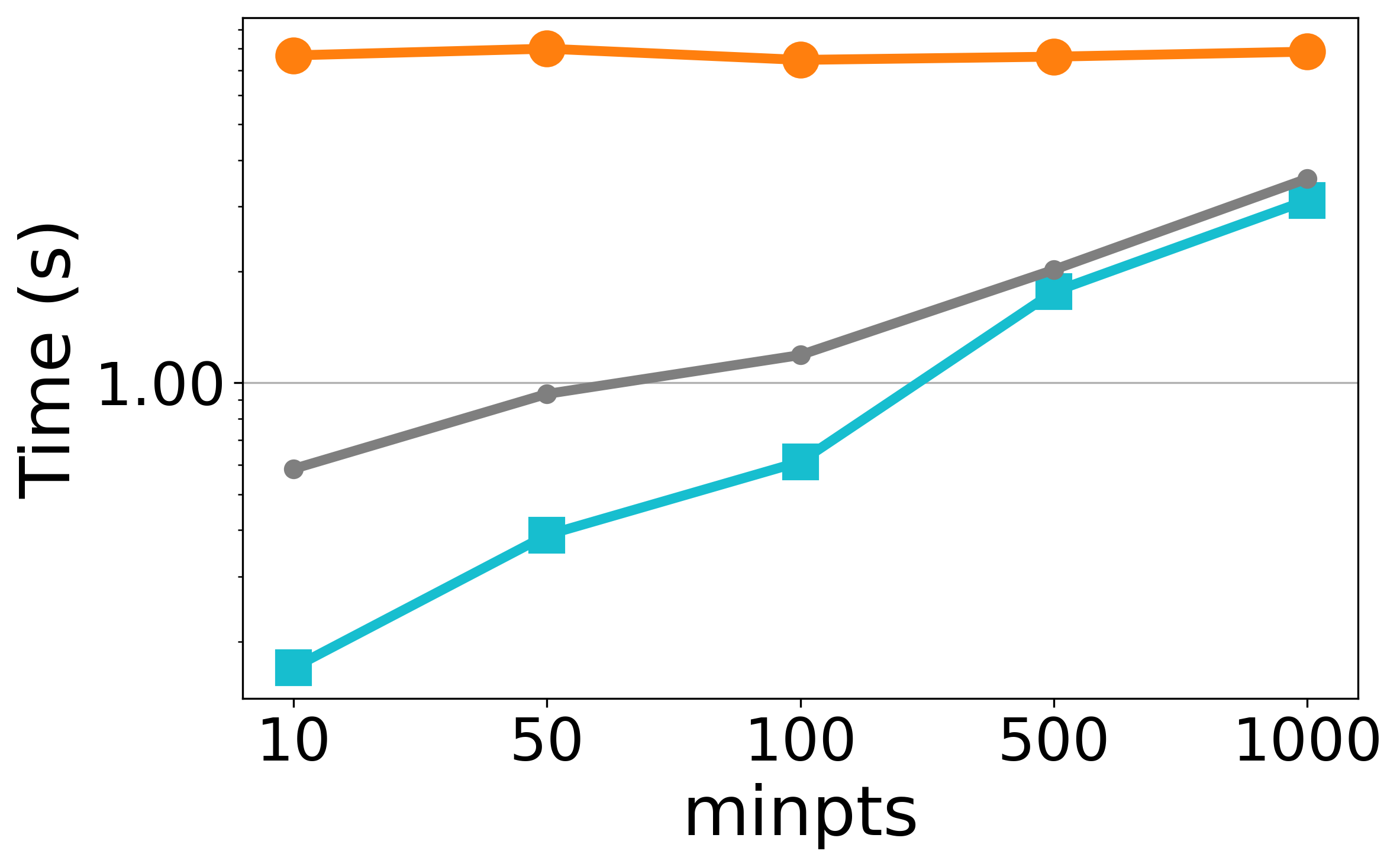}}

    \caption{Impact of the \minpts{} parameter on the execution time.}\label{f:sweep_minpts}
\end{figure*}

\Cref{f:sweep_minpts} shows the effect of varying the \minpts parameter while
keeping $\eps$ and the problem size fixed at the default values.

We observed that in most situations the algorithms exhibit little change in the
behavior, except for \fdbscandense which trends slower for larger \minpts
values as the number of the dense cells decreases. For many datasets, \fdbscan
performs faster than \fdbscandense due to the chosen fixed value of $\eps$. The
growth in \fdbscan results is explained by the longer preprocessing
phase, as the early termination only happens once \minpts neighbors are found;
the main phase is almost unaffected by the \minpts parameter. The preprocessing
phase of \fdbscandense is affected in a similar way, but in addition, the main
phase also takes longer due to larger mixed hierarchy sizes due to lower number
of dense cells. This is particularly noticeable in \dataset{2D-Porto} and
\dataset{7D-Household}. We see that either \fdbscan or \fdbscandense are still
universally the fastest algorithms, often by a large margin.

\subsubsection*{Impact of the number of points in the dataset}

\begin{figure*}[!t]
    \centering
    \includegraphics[width=0.5\textwidth]{figures/sweep_legend.png} \\ [-2ex]

    \subfloat[2D-NGSIM]{\includegraphics[width=0.24\textwidth]{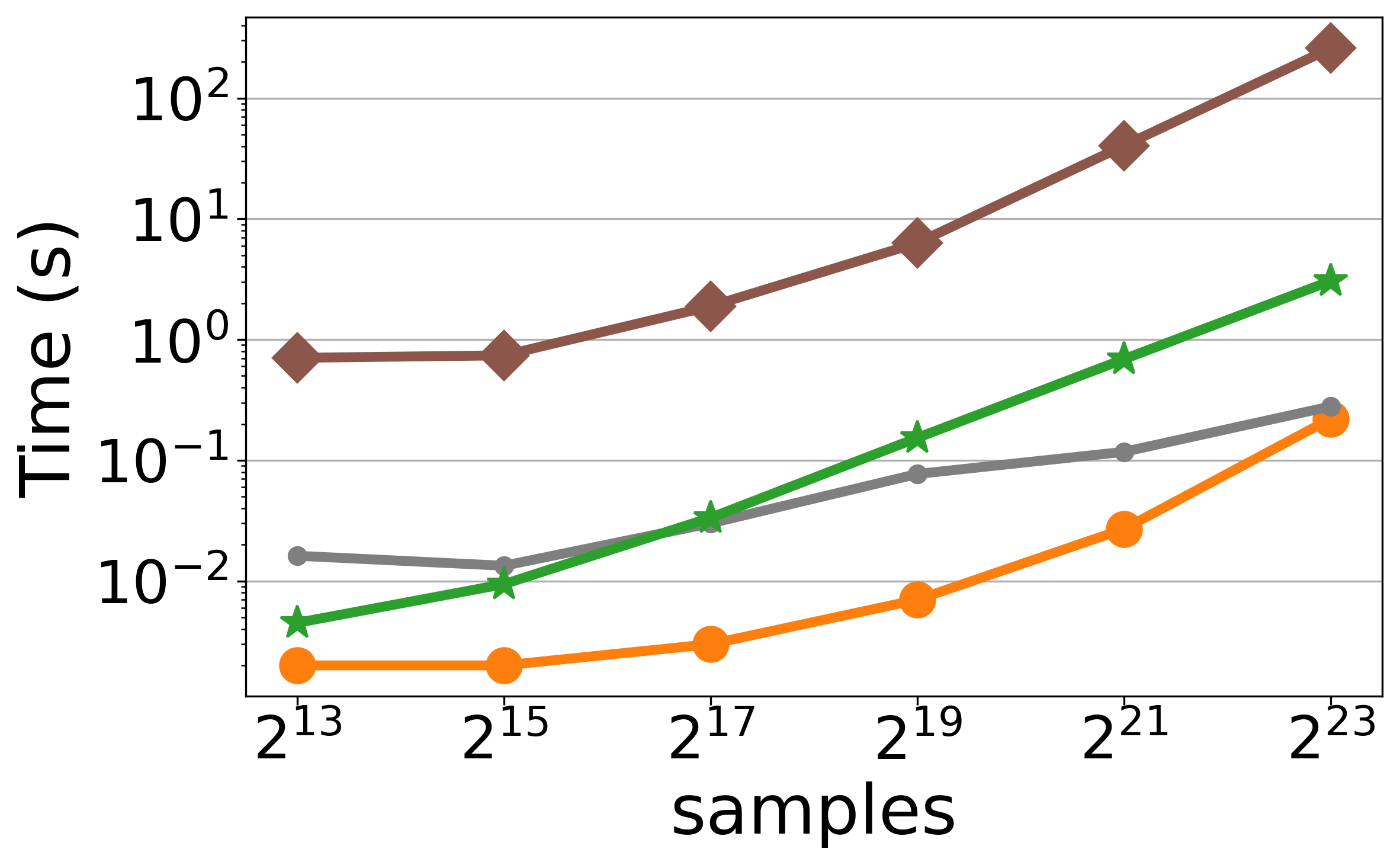}}
    \subfloat[2D-Porto]{\includegraphics[width=0.24\textwidth]{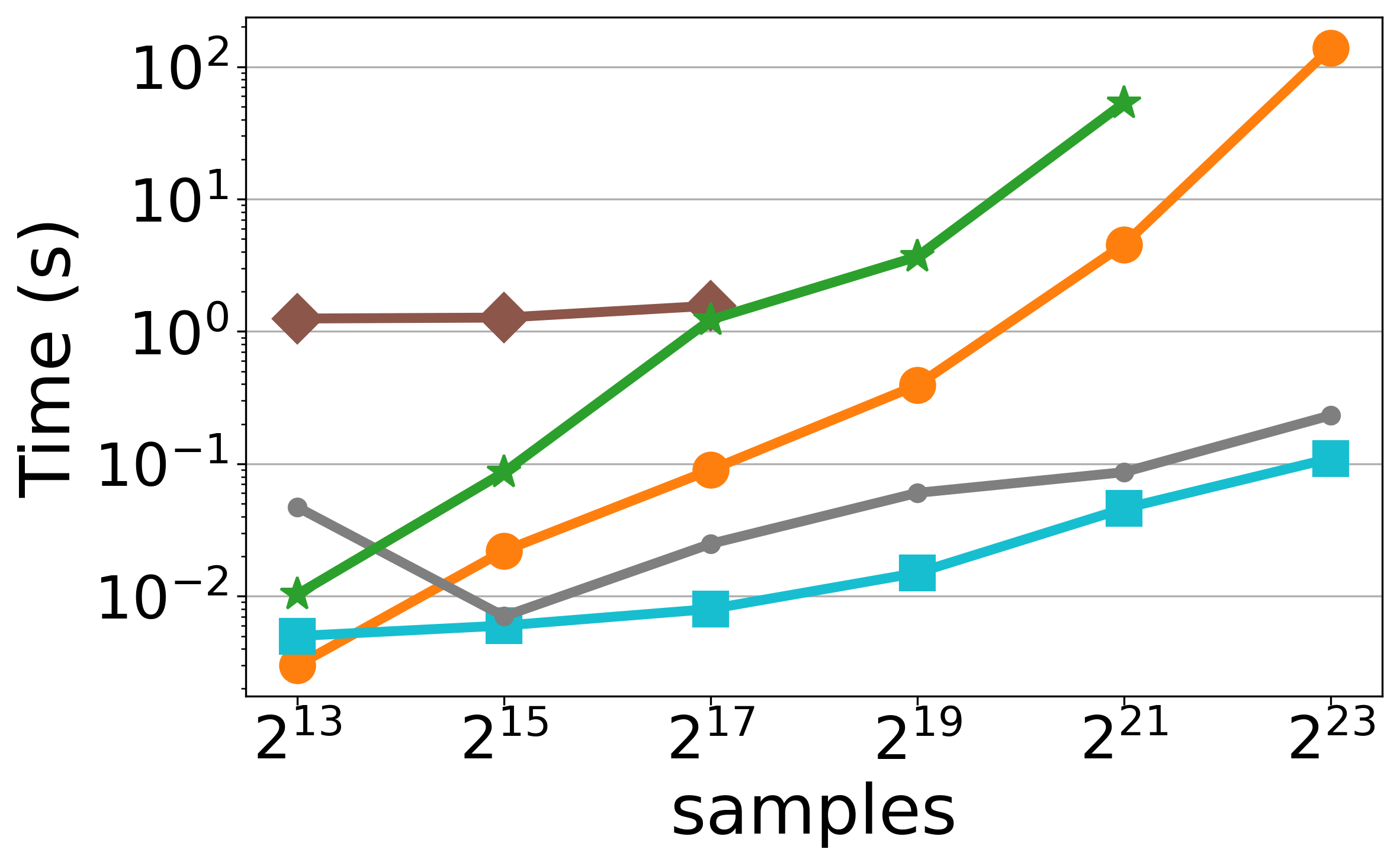}}
    \subfloat[2D-SS-simden]{\includegraphics[width=0.24\textwidth]{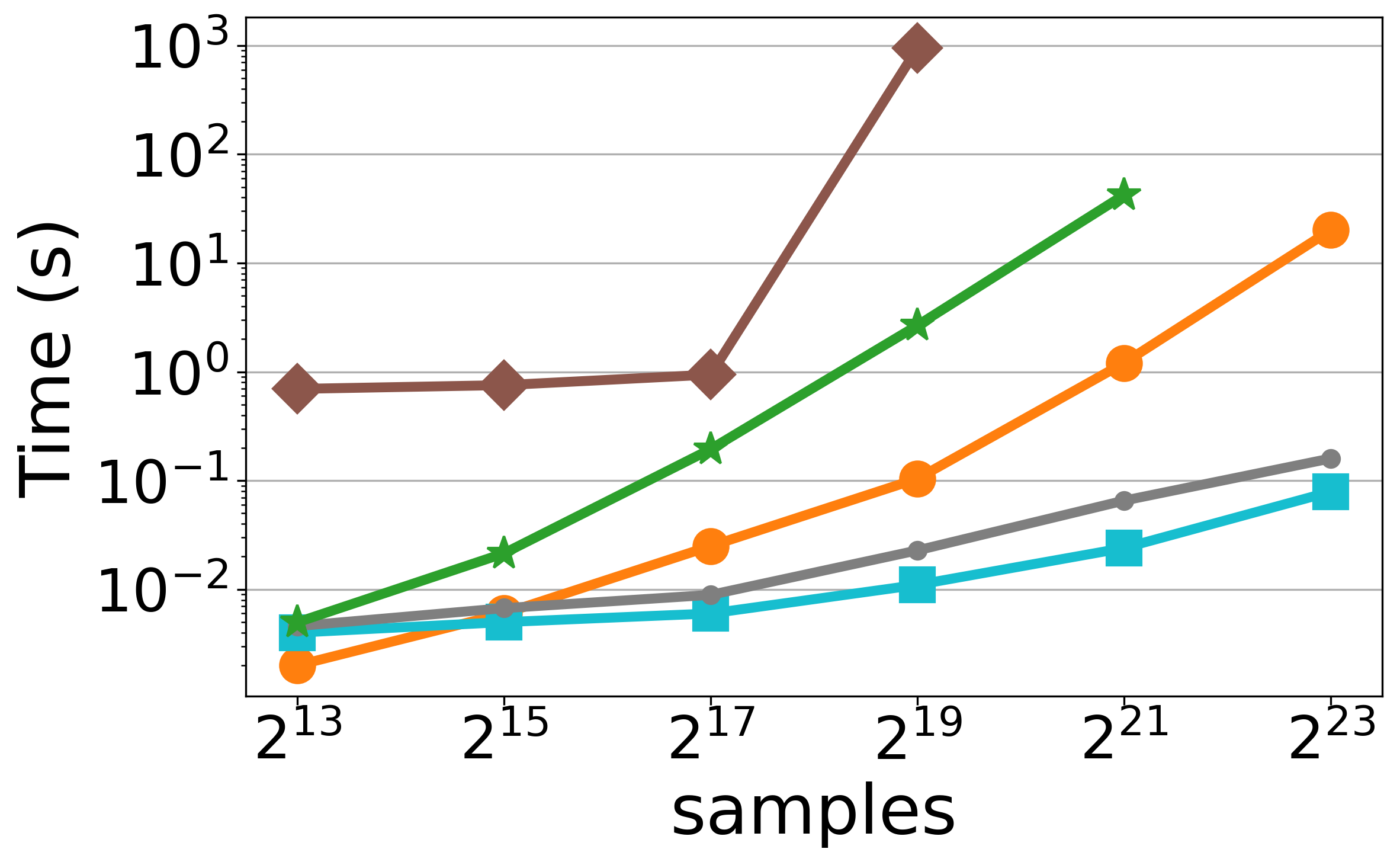}}
    \subfloat[2D-SS-varden]{\includegraphics[width=0.24\textwidth]{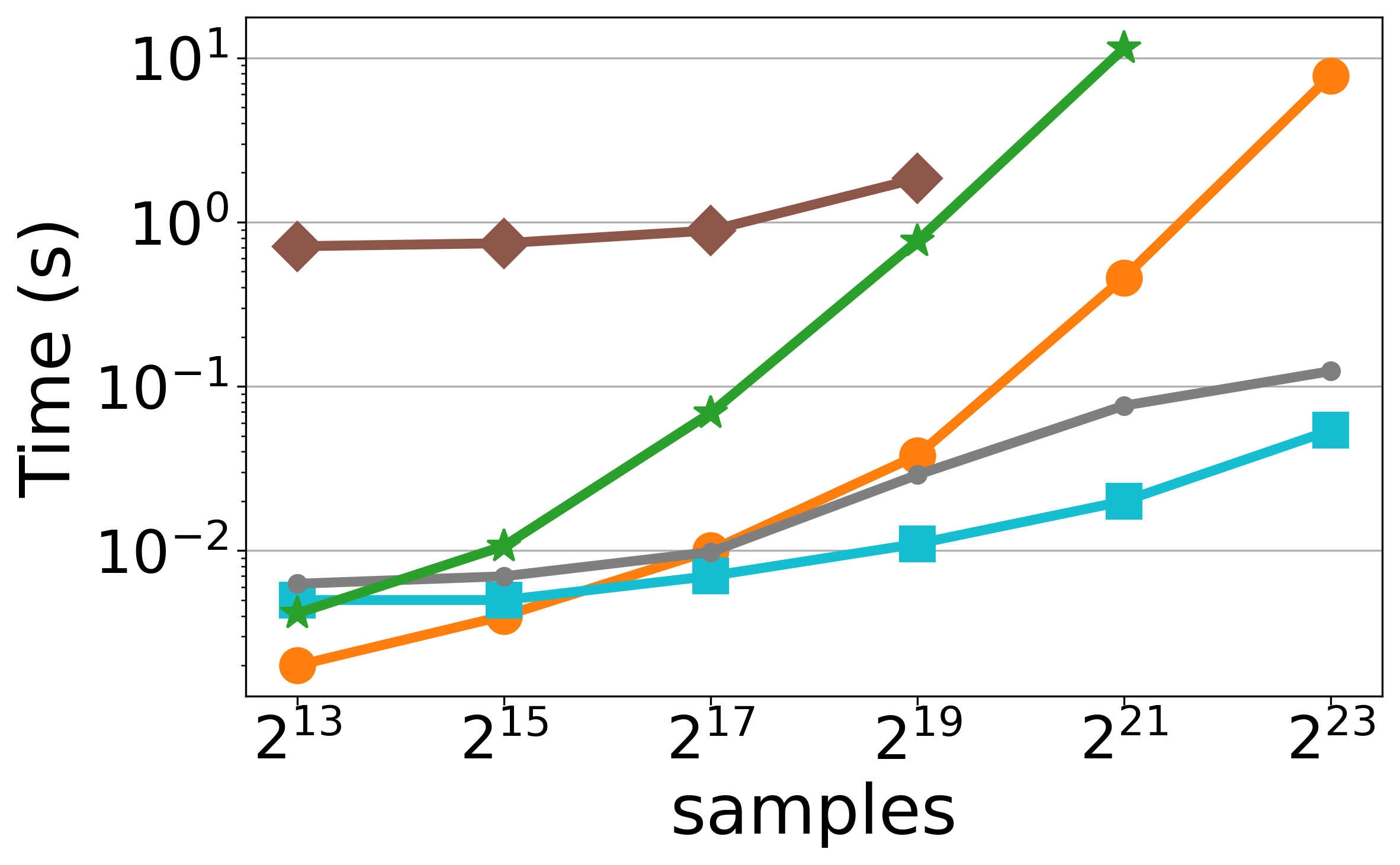}} \\[-2ex]

    \subfloat[3D-Hacc]{\includegraphics[width=0.24\textwidth]{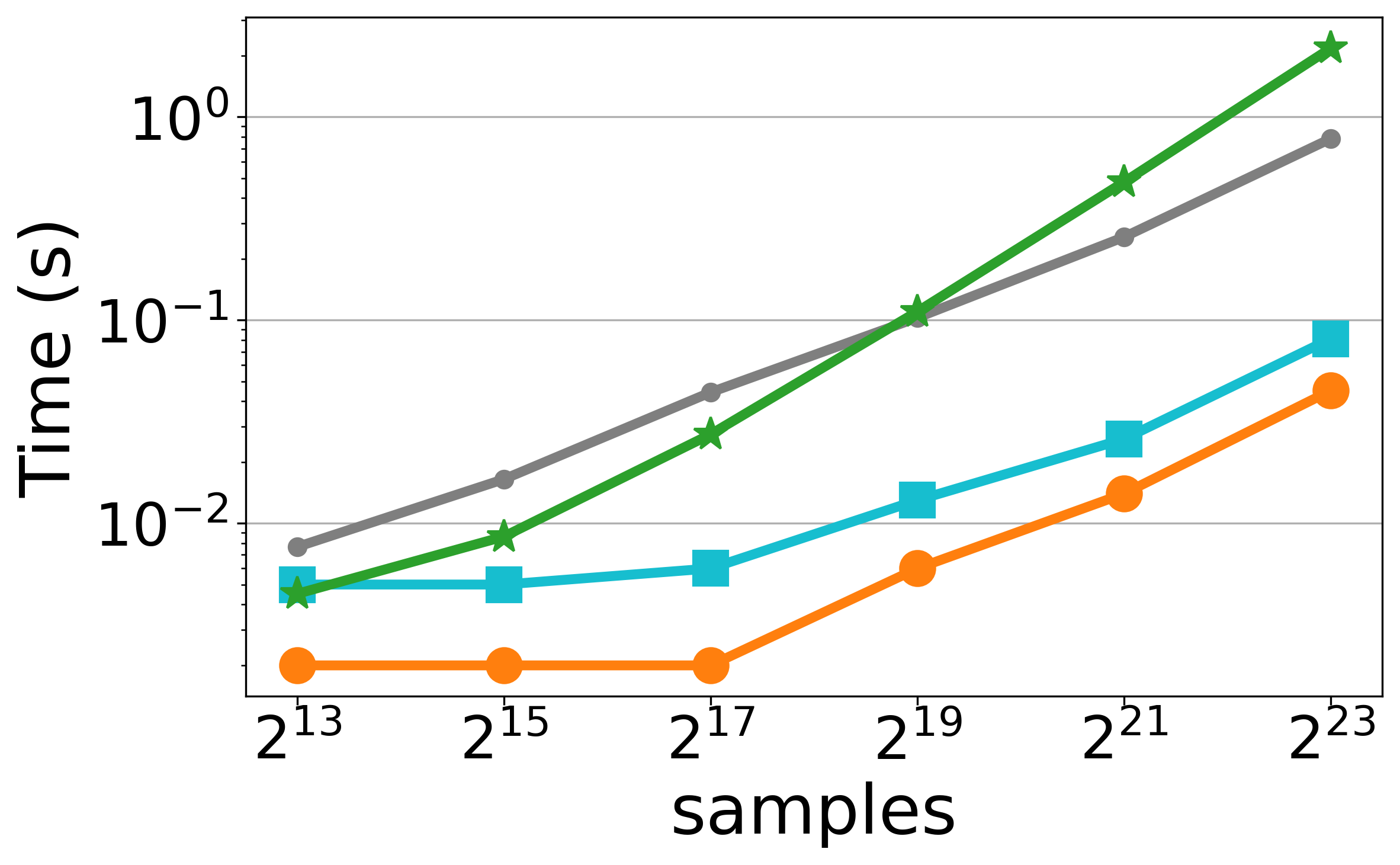}}
    \subfloat[3D-SS-varden]{\includegraphics[width=0.24\textwidth]{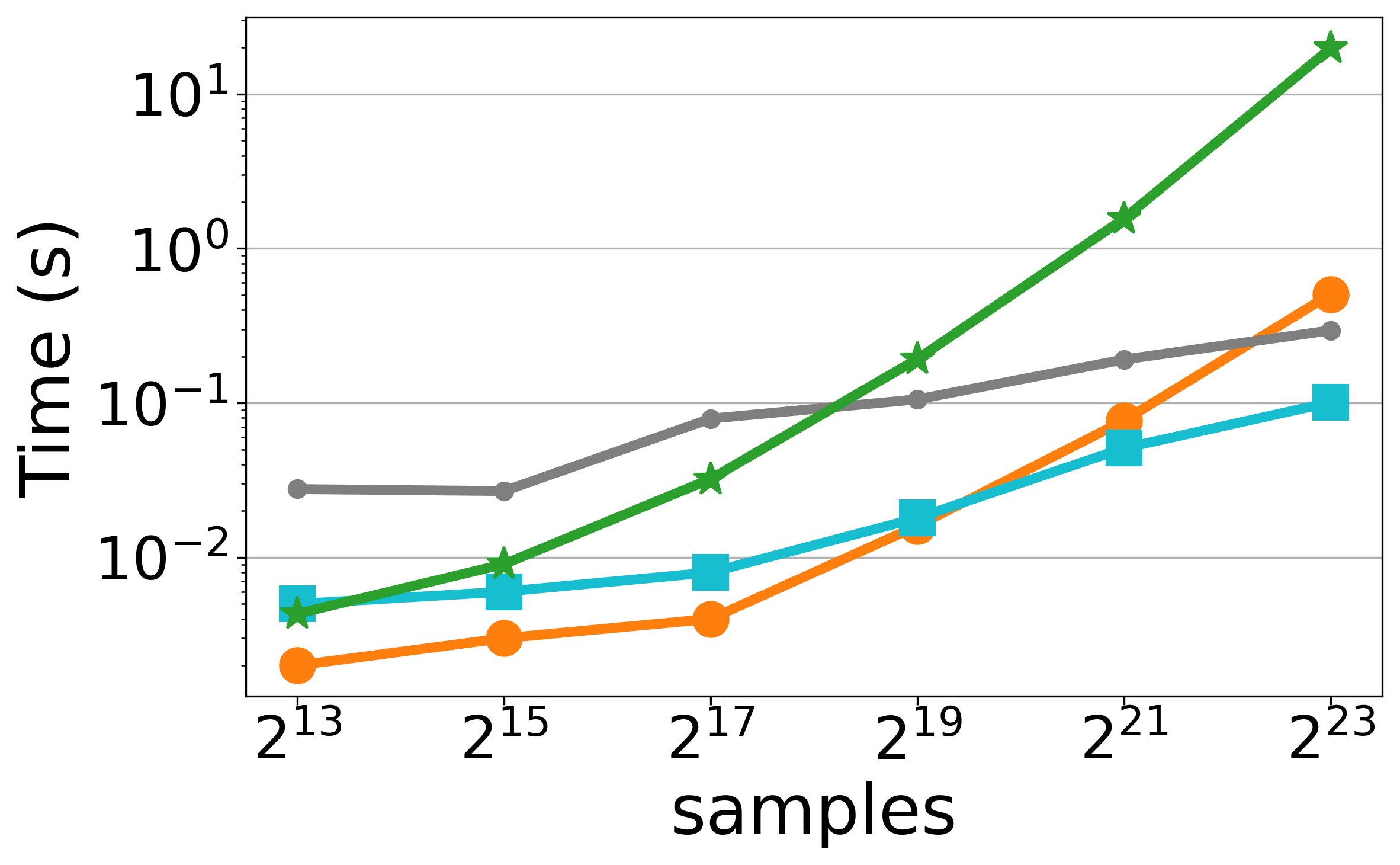}}
    \subfloat[3D-SS-varden]{\includegraphics[width=0.24\textwidth]{figures/3D-ss-var_samples.png}}
    \subfloat[5D-SS-simden]{\includegraphics[width=0.24\textwidth]{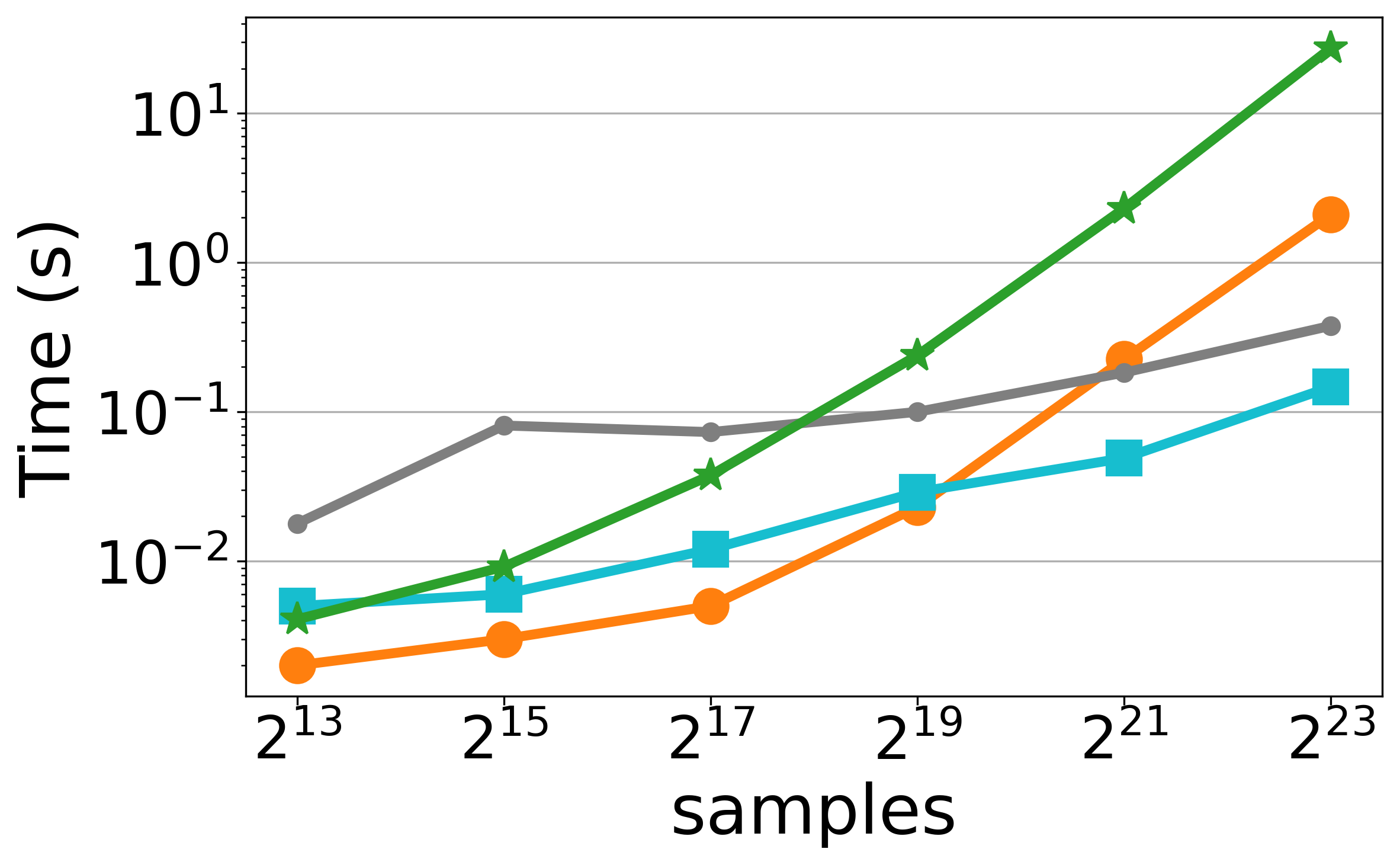}} \\[-2ex]

    \subfloat[5D-SS-varden]{\includegraphics[width=0.24\textwidth]{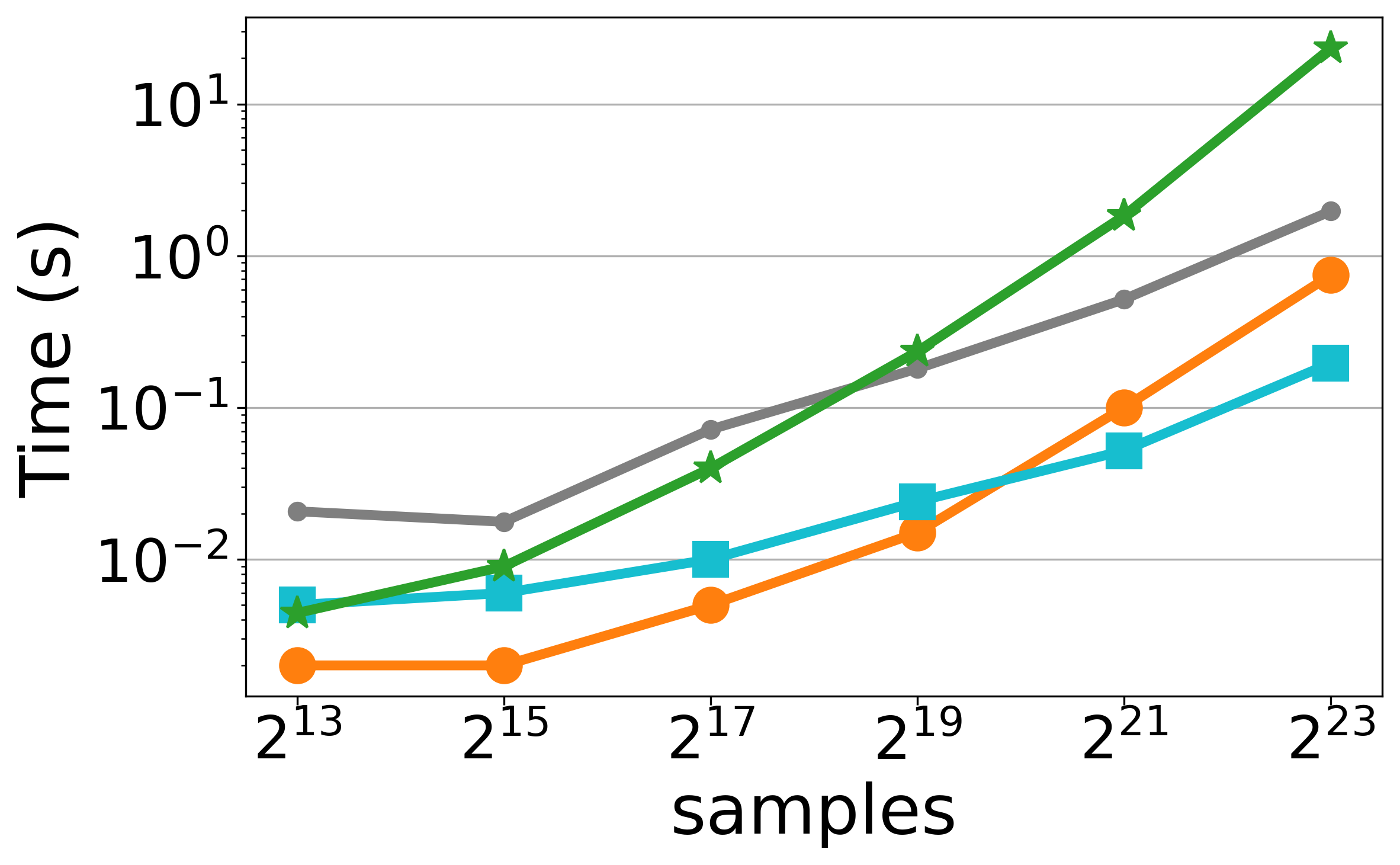}}
    \subfloat[7D-SS-simden]{\includegraphics[width=0.24\textwidth]{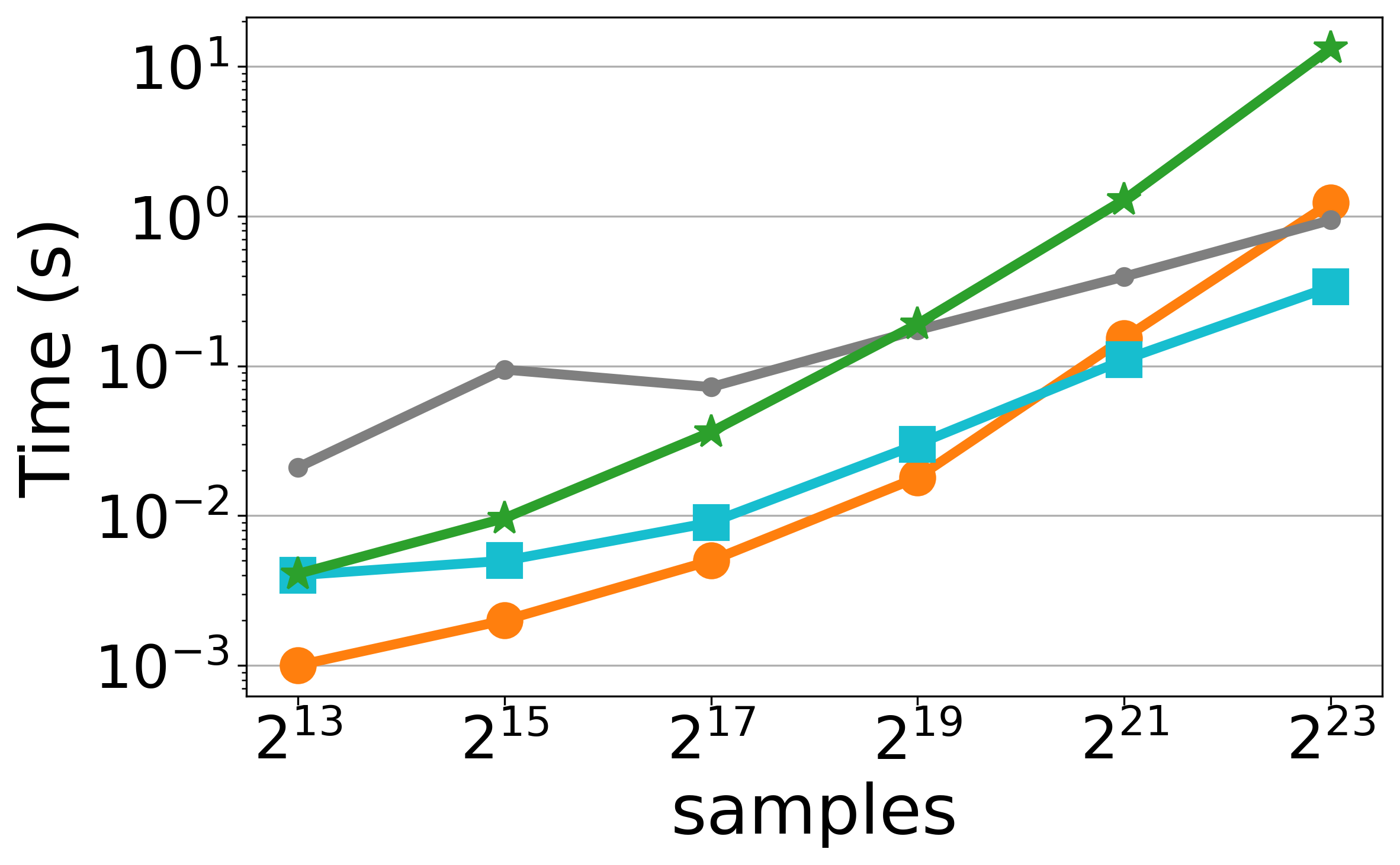}}
    \subfloat[7D-SS-varden]{\includegraphics[width=0.24\textwidth]{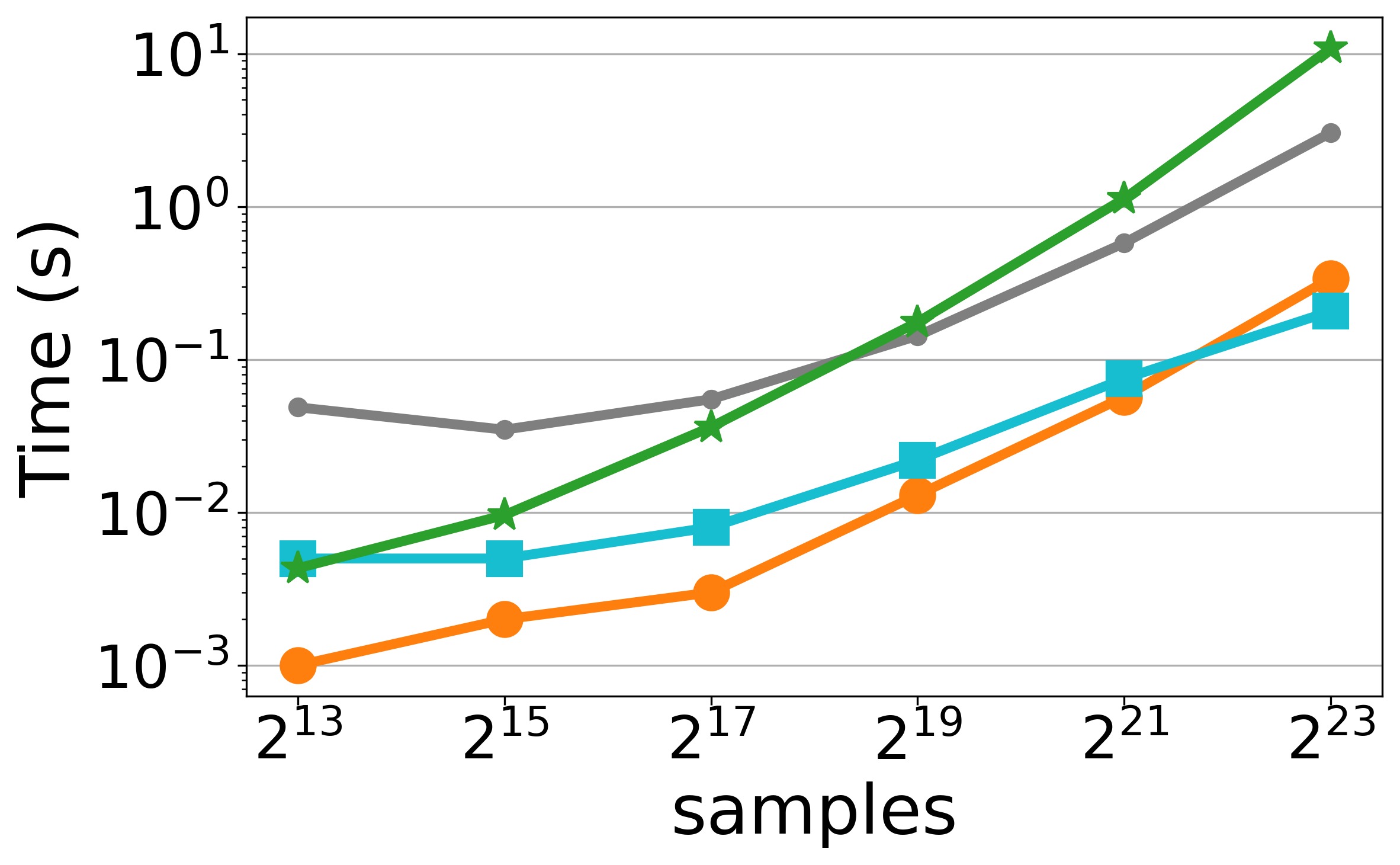}}
    \subfloat[7D-Household]{\includegraphics[width=0.24\textwidth]{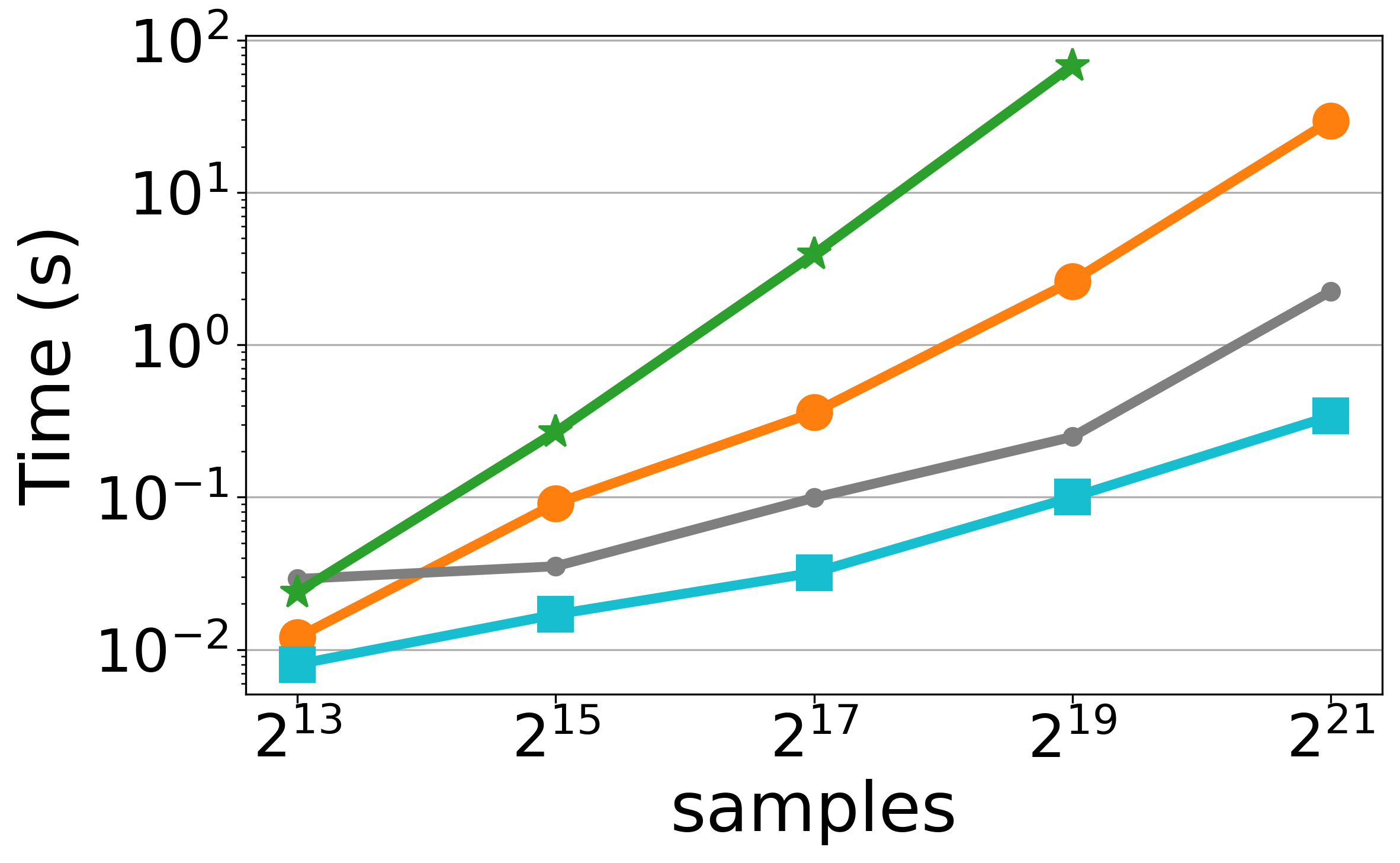}}

    \caption{Impact of the number of samples drawn from a dataset on the execution time.}\label{f:sweep_samples}
\end{figure*}

For our final comparison, we varied the size of the problem by increasing the
number of drawn samples for each dataset while keeping the values of $\eps$ and
\minpts fixed. We chose random sampling as we could not rely on the
organization points in the datasets. However, this results in the problems
becoming denser with increasing size, affecting the performance in addition to
the increases in size.

Figure~\ref{f:sweep_samples} presents the results, shown in log-log scale.
\tepp and \fdbscandense scale similarly and slower than \pdsdbscan and
\fdbscandense. Between \fdbscan and \fdbscandense, for almost all datasets
there is a point at which the \fdbscandense becomes faster due to reaching
sufficient density. Both \gdbscan and \pdsdbscan are clear outliers in terms of
performance.

In addition to missing \fdbscandense data points for the \dataset{2D-NGSIM} due
to the loss of precision, we also note \gdbscan running out of memory at very
modest problem sizes. This is expected as \gdbscan stores the full adjacency
matrix data, so that even 40GB A100 memory is not sufficient for storage.

\subsubsection*{Summary}

\fdbscan and \fdbscandense clearly prove to be very competitive algorithms,
often outperforming other existing algorithms by an order of magnitude, with
\fdbscandense being the typically the much faster of the two. Both algorithms
proposed in this paper do not suffer from the significant memory limitations.
The closest competitor to the algorithms is the \tepp multi-threaded
implementation.

\subsection{Performance portability}\label{s:perf_portability}
\begin{figure*}[!t]
    \centering
    \includegraphics[width=0.85\textwidth]{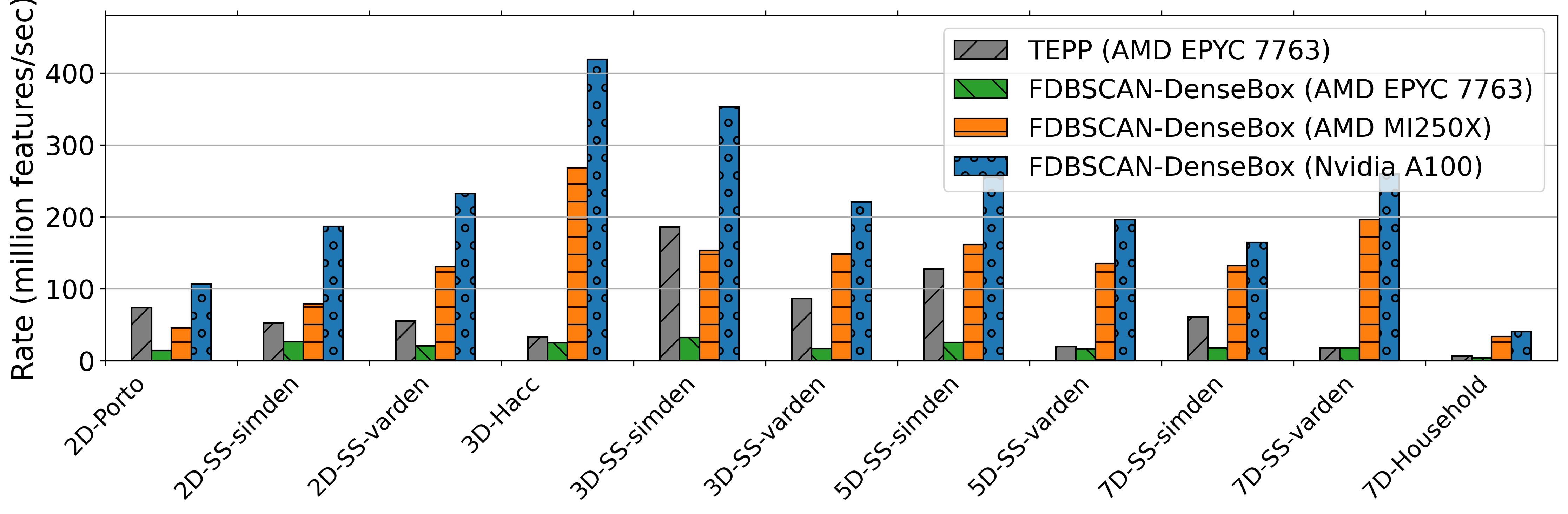}

    \caption{Rate comparison across different hardware architectures.}\label{f:perf_portability}
\end{figure*}

In this Section, we discuss the performance portability of the implemented
algorithms through the use of the Kokkos library~\cite{kokkos2022}.
\Cref{f:perf_portability} shows the performance of the \fdbscandense algorithm
on different hardware: \amdcpu (through OpenMP backend), \amdgpu (through HIP
backend), and \nvidiagpu (through CUDA backend). \tepp baseline is provided for
\amdcpu. The results are presented as the rate, million features (product of
the number of points and dimension) per second.

We see that \amdgpu is 1.2-2.3$\times$ slower than \nvidiagpu, which is
explained by using a single GCD. The OpenMP implementation is 1.0-5.7$\times$
slower than \tepp, and is expected given that the algorithm is designed for GPU
architectures.

Similar performance portability results hold for the \fdbscan algorithm.

\section{Conclusions and future work}\label{s:conclusions}

We presented a general parallel approach for \dbscan on GPUs, and introduced
two algorithms based on a bounding volume hierarchy tree implementation. These
algorithms were evaluated against the other existing CPU and GPU algorithms,
demonstrating their excellent performance. The algorithms were shown to be
performance portable and able to run on a variety of hardware architectures,
including multi-threaded CPUs and GPUs. We showed that a special treatment of
dense areas by using an auxiliary Cartesian grid is advantageous in many
situations.

Algorithmically, we see a number of research directions to pursue.
Similar to~\cite{gowanlock2019a}, we envision using a heuristic to
automatically switch between \fdbscan and \fdbscandense for a given problem.
An introduction of a batched mode is of interest for applications where the
data and the index do not fit in the GPU memory.
Other directions of research include combining the proposed approach with
distributed computations, lowering memory requirements of the used search
index, and incorporating other \dbscan variants such as \dbscans.

\unless\ifjournal
\section*{CRediT author statement}

\textbf{Andrey Prokopenko}: Conceptualization, Investigation, Software, Writing - original draft, Validation.
\newline
\textbf{Damien Lebrun-Grandi\'e}: Software, Writing - review and editing.
\newline
\textbf{Daniel Arndt}: Investigation, Writing - review and editing.
\fi

\unless\ifblind
\ifjournal
\begin{acks}
\else
\section*{Acknowledgements}
\fi
  The authors are grateful to Dr. Eleazar Leal for providing the source code for
  the algorithms used in~\cite{mustafa2019} paper for comparison.
  This research was supported by the Exascale Computing Project (17-SC-20-SC), a
  collaborative effort of the U.S. Department of Energy Office of Science and
  the National Nuclear Security Administration.
  This research used resources of the Oak Ridge Leadership Computing Facility at
  the Oak Ridge National Laboratory, which is supported by the Office of Science
  of the U.S. Department of Energy under Contract No. DE-AC05-00OR22725.
\ifjournal
\end{acks}
\fi
\fi

\ifjournal
  \balance
  \bibliographystyle{ACM-Reference-Format}
\else
  \bibliographystyle{apalike}
\fi
\bibliography{main}

\end{document}